\documentclass[prd,nofootinbib,preprint,superscriptaddress]{revtex4}

\usepackage{amsmath, amssymb, amsthm, graphicx, epsfig, fancyhdr, slashed, ulem}
\usepackage{comment}
\usepackage{tikzsymbols}
\usepackage{natbib}
\usepackage{float}
\usepackage{hyperref}
\usepackage{tikz}
\usepackage{subfigure}

\definecolor{lime}{HTML}{A6CE39}
\DeclareRobustCommand{\orcidicon}{
	\begin{tikzpicture}
	\draw[lime, fill=lime] (0,0) 
	circle [radius=0.2] 
	node[white] {{\fontfamily{qag}\selectfont \tiny ID}};
	\draw[white, fill=white] (-0.0625,0.095) 
	circle [radius=0.007];
	\end{tikzpicture}
	\hspace{-2mm}
}

\foreach \x in {A, ..., Z}{\expandafter\xdef\csname orcid\x\endcsname{\noexpand\href{https://orcid.org/\csname orcidauthor\x\endcsname}
			{\noexpand\orcidicon}}
}

\newcommand{\be}{\begin{equation}}
\newcommand{\ee}{\end{equation}}
\newcommand{\bea}{\begin{eqnarray}}
\newcommand{\eea}{\end{eqnarray}}

\newcommand{\ux}{U\left(1\right)_X}
\newcommand{\xmu}{X_\mu}
\newcommand{\lhs}{\lambda_{HS}}
\newcommand{\ls}{\lambda_S}
\newcommand{\lh}{\lambda_H}
\newcommand{\mpl}{M_\text{pl}}
\begin{document}

\title{Scale Invariant FIMP Miracle}

% \title{Laboratory Signatures of FIMP in Conformal \\ Extensions of the Standard Model}

%\title{Dynamical Generation of Electroweak \\ and Freeze-in Dark Matter Scales}

\author{Basabendu Barman\orcidA{}}
\email{basabendu88barman@gmail.com}
\affiliation{Centro de Investigaciones, Universidad Antonio Nariño\\
Carrera 3 este \# 47A-15, Bogotá, Colombia}

\author{Anish Ghoshal\orcidB{}}
\email{anishghoshal1@protonmail.com}
\affiliation{Institute of Theoretical Physics, Faculty of Physics, University of Warsaw,ul.  Pasteura 5, 02-093 Warsaw, Poland}
\affiliation{INFN - Sezione Roma “Tor Vergata”, Via della Ricerca Scientifica 1, 00133, Rome, Italy}

%%%%%%%%%%%%%
\begin{abstract}
%\textit{We study in a scale invariant model the dark matter scale and the electroweak physics scale is generated dynamically. We focus on the case of Freeze-in DM.}

We study the freeze-in production of vector dark matter (DM) in a classically scale invariant theory, where the Standard Model (SM) is augmented with an abelian $U(1)_X$ gauge symmetry that is spontaneously broken due to the non-zero vacuum expectation value (VEV) of a scalar charged under the $U(1)_X$. Generating the SM Higgs mass at 1-loop level, it leaves only two parameters in the dark sector, namely, the DM mass $m_X$ and the gauge coupling $g_X$ as independent, and supplement with a naturally light dark scalar particle. We show, for $g_X\sim\mathcal{O}\left(10^{-5}\right)$, it is possible to produce the DM X out-of-equilibrium in the early Universe, satisfying the observed relic abundance for $m_X\sim\mathcal{O}\left(\text{TeV}\right)$, which in turn also determines the scalar mixing angle $\sin \theta\sim\mathcal{O}\left(10^{-5}\right)$. The presence of such naturally light scalar mediator with tiny mixing with the SM,
opens up the possibility for the model to be explored in direct search experiment, which otherwise is insensitive to standard freeze-in scenarios. Moreover we show that even with such feeble couplings, necessary for the DM freeze-in, the scenario is testable in several light dark sector searches (e.g., in DUNE and in FASER-II),  
satisfying constraints from the observed relic abundance as well as big bang nucleosynthesis (BBN). Particularly, we find, regions of the parameter space with $m_X$ $\gtrsim 1.8$ TeV are insensitive to direct detection probes but still can become accessible in lifetime frontier searches, courtesy to the underlying scale invariance of the theory.

% direct detection experiments may leave a desert region in the parameter space for DM mass $\gtrsim 1.8$ TeV, where the intensity frontier experiments can be found to be efficient in probing the light scalar.   
\end{abstract}

%%%%%%%%%%%%%%%%
\begin{flushright}
 PI/UAN-2021-698FT
\end{flushright}
%%%%%%%%%%%%%%%
\maketitle 

{
  \hypersetup{linkcolor=black}
  \tableofcontents
}
%%%%%%%%%%%%%%%%%%%%%%%
\section{Introduction}\label{sec:intro}
%%%%%%%%%%%%%%%%%%%%%%%

The discovery of a Higgs-like boson of mass $\sim$ 125 GeV at the Large Hadron Collider~\cite{ATLAS:2012yve, CMS:2012qbp} has provided the last missing piece of the Standard Model (SM) particle physics. Although the Higgs mechanism~\cite{PhysRevLett.13.321, PhysRevLett.13.508, Guralnik:1964eu} satisfactorily provides masses to all the candidates within the SM in a gauge invariant manner, the Higgs itself carries an ad-hoc negative mass term which is provided by hand at the weak scale. As such, it is natural to assume that the SM Higgs mass will be unstabilized against the Planck scale (or any cut-off scale present in the UV) by quadratically diverging quantum corrections unless large fine-tuned cancellation of the associated quadratic divergences is imposed~\cite{PhysRevD.20.2619}. In fact, seeking resolutions to this so called ``naturalness problem" has been the major driving force behind numerous beyond the SM (BSM) extensions, starting from the good old supersymmetry (SUSY), compositeness, extra dimensions (for a review, see Ref.~\cite{Giudice:2017pzm}) \footnote{In context to non-local QFT with higher-derivatives, motivated from string field theories, it has been found that the naturalness issue is somewhat reduced in these theories due to the presence of UV-fixed point \cite{Ghoshal:2017egr,Buoninfante:2018gce,Ghoshal:2018gpq,Ghoshal:2020lfd}.}. In order to stabilize the electroweak scale against the radiative corrections we can promote scale invariance to be a symmetry of the action at the classical level which eliminates the $\mu^2$ mass term or any other dimensionful parameters from the theory\footnote{In a larger context, in BSM scenarios, assuming no scale is fundamental in nature and all mass scales including the EW and Planck scales are generated dynamically can be seen as direction of model-building for the hierarchy problem in the SM~\cite{Adler:1982ri,Coleman:1973jx,Salvio:2014soa,Einhorn:2014gfa,Einhorn:2016mws,Einhorn:2015lzy,Foot:2007iy,AlexanderNunneley:2010nw,Englert:2013gz,Hambye:2013sna,Farzinnia:2013pga,Altmannshofer:2014vra,Holthausen:2013ota,Salvio:2014soa,Einhorn:2014gfa,Kannike:2015apa,Farzinnia:2015fka,Kannike:2016bny}.  In context to cosmology scale invariance provides naturally flat inflationary potentials \cite{Khoze:2013uia,Kannike:2014mia,Rinaldi:2014gha,Salvio:2014soa,Kannike:2015apa,Kannike:2015fom,Barrie:2016rnv,Tambalo:2016eqr}, and also leads to strong first-order phase transitions in early universe and consequently high amplitude detectable gravitational wave (GW) signals due to dominance of thermal corrections in absence of tree-level mass terms~\cite{Jaeckel:2016jlh, Marzola:2017jzl, Iso:2017uuu, Baldes:2018emh, Prokopec:2018tnq, Brdar:2018num, Marzo:2018nov, Ghoshal:2020vud}. }. In the SM one can radiatively generate nonzero Higgs mass and spontaneous electroweak symmetry breaking (EWSB) as pointed out by Coleman and Weinberg (CW) in their seminal work in Ref.~\cite{PhysRevD.7.1888} but it making the Higgs potential unbounded from below at one-loop given the experimentally observed masses of top quark and electroweak gauge bosons. To resolve this problem we thus need to look beyond the minimal SM, where additional bosonic contributions\footnote{Recently the Neutrino Option idea considers threshold corrections (as an alternative to bosonic corrections) to generate the Higgs mass via fermionic loop~\cite{Brivio:2017dfq}.}. can lead to phenomenologically successful models. A simple way to address this issue is to generate the mass scale in a ``hidden sector" and then transmit it to the SM thus avoiding the direct CW relation between the SM gauge boson masses and the mass of the SM Higgs. Such attempts have already been made with or without introducing a suitable dark matter (DM) candidate in Refs.~\cite{Foot:2007as, Iso:2009ss, Iso:2009nw, Okada:2012sg, Farzinnia:2013pga, Englert:2013gz, Hambye:2013dgv, Kang:2014cia, Kang:2015aqa, Hambye:2013sna, Kubo:2014ova, Das:2015nwk, Humbert:2015epa, Humbert:2015yva, Haba:2015lka, Haba:2015yfa, Das:2016zue, Hambye:2018qjv,  Kubo:2018kho,Brdar:2018vjq, YaserAyazi:2019caf, Mohamadnejad:2019vzg, Okada:2019opp, Mohapatra:2019ysk,Okada:2020evk, Mohapatra:2020bze, Gialamas:2021enw, Li:2021pnv}.

The SM also fails to provide a viable DM candidate, whose existence is already been proven beyond any doubt from several astrophysical~\cite{Zwicky:1933gu, Zwicky:1937zza, Rubin:1970zza, Clowe:2006eq} and cosmological~\cite{Hu:2001bc, Aghanim:2018eyx} evidences (for a review, see, e.g. Refs.~\cite{Jungman:1995df, Bertone:2004pz, Feng:2010gw}). Therefore, one has to look beyond the realms of the SM to explain the DM puzzle, if we assume it is of particle origin. There already exists a very attractive scenario known as the weakly interacting massive particle (WIMP), where the DM candidate remains in thermal equilibrium in the early Universe
down to a temperature at which its number density becomes Boltzmann suppressed, and then the DM chemically  decouples from the thermal bath. This is the vanilla {\it freeze-out} mechanism~\cite{Kolb:1990vq}, where the observed DM abundance is set typically by the weak scale interaction strength which popularly goes by the name ``WIMP miracle." 
% Here the observed relic density of DM which was frozen-out, 
% \begin{equation}
% Y_{\rm DM} \equiv \frac{n_{\text{DM}}}{s}  \propto {1\over  \sigma}
% \end{equation}
% where $n_{\rm DM}$ is the DM number density, $s$ the entropy density of the Universe and $\sigma$ is the typical cross section responsible for chemical equilibrium.
Various experimental efforts like the direct detection~\cite{XENON:2018voc, XENON:2020kmp, PandaX:2018wtu} experiments typically search for the recoil of nuclei from collisions with WIMPs from the local halo of DM~\cite{Goodman:1984dc,Ahlen:1987mn}, while attempts have also been made to search for DM signal in collider (see, for example, Refs.~\cite{Penning:2017tmb,Kahlhoefer:2017dnp} and the references therein) experiments. However, no significant excess have been found till date in either of these experiments which can guarantee DM discovery.
%\ag{Cite Ddet experiments..}Recently, the XENON1T collaboration has released its latest data \cite{Aprile:2018dbl}.But no evidence for existence of such particles have been recorded.
The observed DM abundance may have been generated also out of
equilibrium by the so-called {\it freeze-in} mechanism~\cite{McDonald:2001vt, Hall:2009bx, Bernal:2017kxu}. In this scenario, the DM particle
couples to the visible sector very feebly, so that chemical equilibrium is never achieved. The DM particles are then produced by the decay or annihilation of the bath particles, until the production ceased due to cooling of the bath temperature below the relevant mass scale connecting the DM to the visible sector. Typically, freeze-in can further be subdivided into (i) Infra-red (IR), where the DM yield gradually builds up with time and becomes important at lower temperature~\cite{McDonald:2001vt, Hall:2009bx, Chu:2011be, Bernal:2017kxu, Duch:2017khv, Biswas:2018aib, Barman:2019lvm, Heeba:2018wtf}, and (ii) Ultra-violate (UV)~\cite{Hall:2009bx, Elahi:2014fsa, Chen:2017kvz, Bernal:2019mhf, Biswas:2019iqm, Barman:2020plp, Barman:2020ifq, Bernal:2020bfj, Bernal:2020qyu, Barman:2021tgt} where the DM is dominantly produced at very high temperature (as large as the reheating temperature of the Universe) and the yield immediately saturates to a constant value. Because of the tiny coupling strength, the DM particle produced via the freeze-in mechanism are dubbed as the feebly interacting massive particle (FIMP), as opposed to the WIMP. 
% which leads to,
% \begin{equation}
% Y_{\rm DM} \propto  \sigma
% \end{equation} 
% As the DM abundance is typically very tiny,  $Y_{\rm DM}~\approx 4\times 10^{-10}({\rm GeV}/m_{\rm DM})$, the portal couplings
% \footnote{} that are responsible for this processes should also be very tiny and is usually thought to be beyond the reach of typical terrestial or astrophysical searches.
% \ag{Cite some papers where proposals have been made to look for this.}
Even with such tiny portal couplings, in the presence of mediators which connect the visible and the dark sectors, FIMPs may give rise to signatures both in direct search~\cite{Hambye:2018dpi,Heeba:2019jho} and in collider~\cite{YaserAyazi:2015egj, Molinaro:2014lfa, Arcadi:2014dca, Arcadi:2014tsa, Hessler:2016kwm, Belanger:2018sti} experiments. %The appearance of particles with microscopic lifetime can also be a potential signal for freeze-in, that gives rise to several signatures in terms of displaced vertices, stable chraged tracks etc (see, for example, Ref.~\cite{Belanger:2018sti}). 
However, it is important to note here, none of these signatures directly involve the coupling of the DM with the visible sector, that sets the DM relic abundance, but too small to be probed at experiments.

%{\bb{add miracle}}
%as the EW scale is generated at 1-loop and is naturally protected from dangerous quantum corrections from the UV, and

In this work we show that a desirable UV-completion principle in particle theory, like the scale invariance, can on one hand dynamically generate the scales of electroweak and dark matter physics, thereby reducing the naturalness problem of the SM Higgs mass, on the other hand, can also incorporate a testable freeze-in DM scenario. To establish this, here we consider the SM being extended by a $U(1)_X$ gauge symmetry, under which the DM $(X_\mu)$ transforms as an abelian gauge boson. The DM becomes massive once this gauge symmetry is spontaneously broken by the vacuum expectation value (VEV) of a new scalar $S$ that carries a non-zero $U(1)_X$ charge. This gives rise to the typical Higgs portal model, where the DM communicates with the visible sector via the portal coupling, with a strength proportional to the scalar mixing. The DM renders stability as it carries an odd $\mathbb{Z}_2$ charge, while all the SM fields are even under the $\mathbb{Z}_2$. It is important to note that the conformal\footnote{We have used the two terms ``scale" invariance and ``conformal" invariance interchangeably in this paper since they are known to be classically equivalent in any four-dimensional unitary and
renormalizable field theory~\cite{Gross:1970tb, CALLAN197042, Coleman:1970je}.} 
gauge extension to the SM leads to extremely economical model building as one is left with only two independent parameters in the model, one of which gets fixed entirely via relic density requirements, thereby leaving very predictive light scalar singlet hunt in several planned and upcoming intensity and lifetime frontier experiments as a novel and direct probe of freeze-in scenarios. We find, in the present scenario, correct DM abundance is achieved via freeze-in for dark gauge coupling $g_X\sim 10^{-5}$ for dark matter mass $m_X\lesssim$ TeV, ensuring the DM interaction rate always falls below the Hubble expansion rate. The scale invariance also gives rise to a naturally light scalar with mass $m_2\sim\mathcal{O}(\text{MeV})\ll m_X$ . This results in an enhanced recoil rate in DM scattering experiments leading to a detectable DM signal which is in sharp contrast to the usual FIMP paradigm, where DM direct detection is simply impossible due to extreme smallness of the coupling of the DM with the visible sector as mentioned earlier \footnote{The exchange of such a light mediator can also induce large self-interaction, a possibility that is of much interest because it may alleviate the possible shortcomings of collisionless dark matter on small scales (core/cusp~\cite{Spergel:1999mh} and too-big-to-fail~\cite{Kaplinghat:2013yxa} problems, see~\cite{Tulin:2017ara} for a review).}.  

The crucial outcome of our analysis is the experimental testability of the FIMP model in low energy intensity frontier experiments, which typically search for light dark sector particles, and are capable of probing extremely small mixing angle. This is facilitated by the presence of the light scalar in this model. We find, in order to satisfy bounds from relic density, spin-independent direct detection and big bang nucleosynthesis (BBN), the mixing within the scalar sector has to be $\sin\theta\sim g_X\,\sim\mathcal{O}(10^{-5})$. Such a mixing angle is indeed within the reach of experiments like DUNE~\cite{DUNE:2015lol, Berryman:2019dme},  FASER-II~\cite{Feng:2017vli, FASER:2018eoc, FASER:2018bac, FASER:2019aik}, PS191~\cite{Bernardi:1985ny, Gorbunov:2021ccu}, DarkQuest-Phase2~\cite{Batell:2020vqn}, MATHUSLA~\cite{Curtin:2018mvb} and SHiP~\cite{SHiP:2015vad}.  Most importantly, since the mixing $\theta$ is no more a free parameter (due to the scale invariance of the theory), it rather can be parametrized in terms of $m_X$ and $g_X$, hence, the new gauge coupling that determines the freeze-in abundance, is directly being probed at plethora of current and upcoming experimental facilities. To the best of our knowledge, this has not been realized before in the context of freeze-in DM. Particularly, we find out that scale invariance {\it uniquely} dictates the mixing $\sin\theta\sim\mathcal{O}\left(10^{-5}\right)$ that  predicts potential discovery at the laboratory experiments. This leads us to dub the scenario as the {\it ``Scale Invariant FIMP Miracle."}
%{\sout {although scale invariance allows only two independent parameters for this model, which conspire among themselves to give rise to a viable DM parameter space,where the DM abundance $\Omega_X\,h^2\propto\left(g_X/m_X\right)^4$, that predicts potential signatures at the laboratory experiments, is a {\it miracle} in itself. Hence we dub the scenario as the {\it ``Scale Invariant FIMP Miracle."}}}
%\ag{\textbf{Can we show some analytic relation here supplementing the text you have written ?}}

%Thus, this model is an example where the freeze-in coupling is directly being probed both at the direct search and in the collider experiments.  

The paper is organized as follows. In Sec.~\ref{sec:model} we have explained the underlying model, elaborating its scale invariant features. The Renormalization Group Equation (RGE) running of various couplings appearing in the theory are discussed under subsection~\ref{sec:rge}. We then move on to the detailed discussion of the DM parameter space in Sec.~\ref{sec:dm-pheno}, where the details of freeze-in yield are mentioned in subsection~\ref{sec:freeze-in}, together with the illustration of available parameter space in subsection~\ref{sec:dm-paramspace}. In Sec.~\ref{sec:expts} we show the reach of several experiments in probing the DM parameter space. Finally, in Sec.~\ref{sec:concl} we conclude with an engaging discussion on our analysis.

%\medskip 

%%%%%%%%%%%%%%%%%%%%%%%%%%%%
\section{The Scale-Invariant Model}
\label{sec:model}
%%%%%%%%%%%%%%%%%%%%%%%%%%%%%
\subsubsection{Symmetries and particle content}
%%%%%%%%%

We consider a dark $\ux$ gauge symmetry and a complex scalar $S$ which is neutral under the SM gauge group but has unit charge under $\ux$\footnote{The phenomenology of abelian vector boson DM has been widely studied in the literature both in the context of freeze-out and freeze-in production without considering scale invariance (see, for example, Refs.~\cite{Servant:2002aq, Birkedal:2006fz, Farzan:2012hh, Baek:2012se, Bian:2013wna, Choi:2013qra, Baek:2013dwa, Baek:2013fsa, Baek:2014goa, Ko:2014gha, Baek:2014poa, Ko:2014loa, Duch:2015jta, Beniwal:2015sdl, Kamon:2017yfx, Duch:2017nbe, Duch:2017khv, Arcadi:2017jqd, Baek:2018aru, Barman:2020ifq}). In presence of scale invariance the phenomenology of WIMP-like vector DM has also been studied~\cite{Hambye:2013dgv,Carone:2013wla, Khoze:2014xha, YaserAyazi:2015egj, Karam:2015jta, Karam:2016rsz, Khoze:2016zfi, Baldes:2018emh, YaserAyazi:2019caf}, while FIMP case has been discussed in~\cite{Heikinheimo:2017ofk, Okada:2020evk}.}. Furthermore we impose a discrete symmetry $\mathbb{Z}_2$, under which all the SM fields transform as even while the new fields transform as odd, such that

\begin{equation}\begin{split}
& X_\mu\to-X_\mu;~~S\to S^\star.                 
                \end{split} 
\end{equation}

\noindent. This forbids the gauge kinetic mixing with the SM $U(1)_Y$ gauge boson $B_\mu$ ensuring the stability of $\xmu$ vector which is our dark matter (DM) candidate\footnote{In the absence of the $\mathbb{Z}_2$ symmetry, the dark gauge boson can still account for all of the DM relic abundance if the kinetic mixing is of the order $\epsilon \lesssim \mathcal{O}\left(10^{-8}\right)$ for DM masses below twice the electron mass~\cite{Arias:2012az, Caputo:2021eaa}, else the cosmological stability condition requires even smaller values $(\lesssim 10^{-15})$ of the kinetic mixing parameter~\cite{Bloch:2016sjj, Lin:2019uvt}.}. The relevant part of the total Lagrangian is given by

\begin{equation}
\mathcal{L}\supset -\frac{1}{4}X_{\mu\nu}X^{\mu\nu}+\left|D_\mu S\right|^2-V\left(H,S\right) 
\end{equation}

\noindent where $H$ is the $SU(2)_L$ scalar doublet and 

\begin{equation}
D_\mu = \partial_\mu + ig_XX_\mu 
\end{equation}

\noindent is the covariant derivative of $S$ with $g_X$ as the new gauge coupling. Since we assume classical scale invariance and renormalizibility,
the only terms allowed in the total effective scalar potential

\begin{equation}
V\left(H,S\right)_\text{cl}=\lambda_H\left(H^\dagger H\right)^2+\lambda_S\left|S\right|^4-\lambda_{HS}\left(H^\dagger H\right)\left|S\right|^2. 
\label{eq:tree-pot}
\end{equation}

The dark and the visible sectors communicate via the renormalizable $\lambda_{HS}$ coupling with the limit $\lambda_{HS}\to 0$  corresponds to the case where the hidden
sector completely decoupled from the visible sector. 

%Note that, we have a completely scale-free potential even in presence of the Higgs portal coupling. We employ the Coleman-Weinberg (CW) mechanism~\cite{PhysRevD.7.1888} which, in the first step, generates a vacuum expectation value (VEV) $\langle S\rangle$ in the hidden sector through dimensional transmutation, which is then transmitted to the Standard Model (SM) via the non-zero Higgs portal. 

%%%%%%%%%%%%%%%%
\subsubsection{Minima of the scalar potential}
%%%%%%%%%%%%%%%%%

The stability of the potential can be verified by completing the square

\begin{equation}\begin{split}
& V\left(H,S\right)=\lambda_H\Bigl(H^\dagger H-\frac{\lambda_{HS}}{2\lambda_H}\left|S\right|^2\Bigr)^2+\left|S\right|^4\Bigl(\lambda_S-\frac{\lambda_{HS}^2}{4\lambda_H}\Bigr)                
                \end{split}
\end{equation}

\noindent which shows the potential is stable as long as

\begin{equation}
4\lambda_H\lambda_S-\lambda_{HS}^2>0. 
\label{eq:stable}
\end{equation}. 

Now, in the unitary gauge, the imaginary component of $S$ can be absorbed as the longitudinal component of $X_\mu$. The tree level potential looks

\begin{equation}\begin{aligned}
V_\text{tree}=\frac{\lh}{4}h^4-\frac{\lhs}{4}h^2\,s^2+\frac{\ls}{4}s^4
\end{aligned}
\label{eq:v2}
\end{equation}

\noindent where we write

\begin{equation}
S = \frac{s}{\sqrt{2}} \,,H=\Bigl\{0,\frac{h}{\sqrt{2}}\Bigr\}^T.    
\end{equation}

\noindent We now construct the Hessian matrix as:

\begin{equation}
\mathcal{H}\equiv\frac{\partial^2 V_\text{cl}}{\partial h_i\partial h_j}.
\end{equation}

\noindent Then, the necessary and sufficient conditions for local minimum of $V_\text{cl}$, corresponding to the vacuum:  $\langle S\rangle=v_s\,,\langle H\rangle=v_h$ are

\begin{equation}\begin{aligned}
&\frac{\partial V_\text{cl}}{\partial h_i}\Big|_{v_s,v_h}=0
\\&
\frac{\partial^2 V_\text{cl}}{\partial^2 h_i}\Big|_{v_s,v_h}>0
\\&
\text{det}\Bigl(\mathcal{H}\left(v_s,v_h\right)\Bigr)>0,
\end{aligned}
\label{eq:cond-min}
\end{equation}

\noindent where `det' stands for the determinant of the Hessian matrix. These  conditions give rise to, for non-vanishing vacuum expectation values (VEVs) %\ag{defined VEV above ?}: 
$$\lambda_H\,\lambda_s = \frac{1}{4}\lambda_{HS}^2$$ and we obtain a relation between the VEVs\footnote{This relation establishes the phenomenon of dimensional transmutation~\cite{Foot:2007iy}.}:

\begin{equation}
\frac{v_h^2}{v_s^2}=\frac{\lambda_{HS}}{2\lambda_H}\label{eq:vev}    
\end{equation}

\noindent which defines a ``flat direction", along which $V_\text{cl}=0$~\cite{Gildener:1976ih, Foot:2007as, Foot:2007iy, Alexander-Nunneley:2010tyr, Farzinnia:2013pga, Chataignier:2018kay, Mohamadnejad:2019vzg}. Further including one-loop corrections will lift this flat direction and generate the true physical vacuum (corresponding to the spontaneous breaking of
classical scale invariance). Since the classical potential is zero along that direction, the one-loop correction necessarily dominates there. Now, the first two relations in Eq.~\eqref{eq:cond-min} require 

%For $g_X\lesssim10^{-4}$, which is required for freeze-in, we find $\lambda_{HS}\lesssim 10^{-10}$ for $m_X\sim 1$ TeV.

\begin{equation}
\lambda_H>0\,,\lambda_s>0\,,\lambda_{HS}>0,    
\end{equation}

\noindent which also satisfies Eq.~\eqref{eq:stable}. However, the determinant of the Hessian turns out to be zero, making the second derivative test inconclusive, and the point $\{v_s,v_h\}$ could be any of a minimum, maximum or saddle point. We find, since at the minimum of the 1-loop effective potential $V_\text{cl}\geq 0$ and $V_\text{eff}^\text{1-loop}<0$, the minimum of $V_\text{eff}^\text{1-loop}$ along the flat direction (given by Eq.~\eqref{eq:vev}), where $V_\text{tree}=0$, is a global minimum of the full potential~\cite{YaserAyazi:2019caf,Mohamadnejad:2019vzg}. Therefore, spontaneous symmetry breaking (SSB) indeed occurs and we expand the fields around the minima

\begin{equation}\begin{split}
& S=\frac{1}{\sqrt{2}}\left(s+v_s\right),~~H=\frac{1}{\sqrt{2}}\Big\{0,h+v_h\Big\}^T.                 
                \end{split}\label{eq:vev-exp}
\end{equation}

As $S$ receives a non-vanishing VEV, the mass of the $\ux$ gauge boson i.e., the DM can be expressed as 

\begin{equation}\begin{aligned}
& m_X =g_X v_s\equiv g_X\,\sqrt{2\lambda_H/\lambda_{HS}}\,v_h
%\implies\lambda_{HS}\lesssim \left(10^{-1}g_X\right)^2.
\end{aligned}\label{eq:mx}
\end{equation}

\noindent which implies, in order to have $m_X\sim\mathcal{O}\left(\text{TeV}\right)$ with $g_X\sim\mathcal{O}\left(10^{-5}\right)$, which will be required for freeze-in, one needs $v_s\gg v_h\equiv v_\text{EW}\simeq 246$ GeV. 

%%%%%%%%%%%%%%%%
\subsubsection{The scalar sector}
%%%%%%%%%%%%%%%%

To calculate the tree-level masses, we expand the potential in Eq.~\eqref{eq:v2} around the vacuum (Eq.~\eqref{eq:vev-exp}), and construct the mass matrix in the weak basis:

\begin{equation}
\mathcal{M}^2=\left(\begin{array}{cc}
 2 v_h^2 \lambda_H & -v_h v_s \lambda_{HS} \\
 -v_h v_s \lambda_{HS} & 2 v_s^2 \lambda_s\\
\end{array}\right).
\end{equation}

\noindent We can then rotate it to the physical (mass) the basis via

\begin{equation}
\begin{pmatrix}
h_1 \\ h_2 
\end{pmatrix} = \begin{pmatrix}
\cos\theta & -\sin\theta\\\sin\theta & \cos\theta
\end{pmatrix}\,\begin{pmatrix}
h \\ s
\end{pmatrix}
\end{equation}

\noindent where the mixing angle is given by

\begin{equation}
\tan2\theta = \frac{2v_h\,v_s}{v_h^2-v_s^2}\implies\sin\theta=\frac{v_h}{\sqrt{v_h^2+v_s^2}}\approx\frac{v_h}{v_s} 
\label{eq:mixing}
\end{equation}

\noindent for $v_s\gg v_h$. In tree-level, the field $h_2$ being along the flat direction, is massless\footnote{This field with vanishing zeroth-order mass, is dubbed as ``scalon"~\cite{Gildener:1976ih}.}. This can be readily seen by obtaining the mass eigenvalues of the diagonalized mass matrix along the flat direction using the mixing angle Eq.~\eqref{eq:mixing}:

\begin{equation}
m_{h_1}^2 = v_h^2\, \left(2\lh+\lhs\right)\,,m_{h_2}^2=0
\end{equation}

\noindent where we note, in the limit $\lhs\to0$ we recover the standard electroweak Higgs state. The massless scalar, on the other hand, will acquire its radiative mass along the flat direction à la Coleman-Weinberg~\cite{PhysRevD.7.1888}, and thus becomes a pesudo-Nambu-Goldstone boson (pNGB) at quantum level. The field $h_1$, on the other hand, is perpendicular to the flat direction which we identify as the SM-like Higgs
% \footnote{If we had $\lhs<0$, then the usual electroweak Higgs would have been the pseudo Goldstone boson~\cite{Foot:2007iy}.} 
observed at the LHC with a mass of $m_{h_1}=125$ GeV~\cite{Mohamadnejad:2019vzg}. After spontaneous symmetry breaking, we can express the couplings of the scalar potential in terms of physical masses and mixings as 

\begin{equation}
\lambda_H= \frac{m_{h_1}^2}{2\,v_h^2}\cos^2\theta\,,\lambda_{HS}=-\frac{m_{h_1}^2}{2v_h\,v_s}\sin2\theta\,,\lambda_s=\frac{m_{h_1}^2}{2v_s^2}\sin^2\theta
\label{eq:cplngs}
\end{equation}

\noindent while the VEV $v_s$ can be expressed as $v_s=m_X/g_X$ following Eq.~\eqref{eq:mx}. Note that, the scale invariance helps to choose minimum number of parameters, and we choose $\{m_X,g_X\}$ as the free parameters for further analysis. All other relevant parameters appearing in the theory can be recasted in terms of $m_X$ and $g_X$:

\begin{equation}\begin{aligned}
& \lh=\frac{m_{h_1}^2}{2v_h^2\left[1+\left(g_X/m_X\right)^2\,v_h^2\right]}\,,\ls=\frac{\left(g_X/m_X\right)^4\,m_{h_1}^2 v_h^2}{2\left[1+\left(g_X/m_X\right)^2\,v_h^2\right]}\,,
\lhs = \frac{g_X^2 \left(m_{h_1}/m_X\right)^2 }{1+\left(g_X/m_X\right)^2\,v_h^2}.
\end{aligned}\label{eq:cplngs2}
\end{equation}

\noindent We again remind the reader about the fact that all the renormalizable couplings in the scalar potential are expressed in terms of the two free parameters of the theory, thanks to the underlying classical scale invariance.   

%%%%%%%%%%%%%%%%
\subsubsection{1-loop effective potential}
%%%%%%%%%%%%%%%%

The scale invariance of the theory gives rise to massless scalar field in the classical level. One loop correction then breaks the scale invariance giving mass to the massless eigenstate $h_2$. Following the approach of Gildener and Weinberg~\cite{Gildener:1976ih}, the 1-loop effetive potential can be approximately written as~\cite{Foot:2007as, Foot:2007iy,Farzinnia:2013pga,YaserAyazi:2019caf,Mohamadnejad:2019vzg, Kannike:2020ppf}

\begin{equation}
V_\text{eff}^\text{1-loop} = \alpha h_2^4+\beta h_2^4\log\frac{h_2^2}{\mu^2}
\label{eq:1-loop}
\end{equation}

\noindent with $\alpha,\beta$ as dimensionless constants

\begin{equation}
\alpha=\frac{1}{64\pi^2\,v^4}\sum_j g_j m_j^4\log\frac{m_j^2}{v^2}\,,\beta=\frac{1}{64\pi^2\,v^4}\sum_j\,g_j\,m_j^4,   
\end{equation}

\noindent where $g_j$ and $m_j$ are the tree-level mass and the internal degrees of freedom of the $j^\text{th}$ particle, $v^2=v_s^2+v_h^2$ and $\mu$ is the renormalization scale. Minimizing Eq.~\eqref{eq:1-loop} we find that the potential has a non-trivial stationary point at

\begin{equation}
\mu = v\,\exp\Biggl(\frac{\alpha}{2\beta}+\frac{1}{4}\Biggr).
\label{eq:1-loop-min}
\end{equation}

\noindent Note that, this shows, the scale of the symmetry breaking is set by the renormalization scale $\mu$. The stationary point is a minimum as long as $\beta$ is positive definite. The 1-loop potential can now be re-written utilizing Eq.~\eqref{eq:1-loop-min} as

\begin{equation}
V_\text{eff}^\text{1-loop} = \beta\,h_2^4\Biggl[\log\frac{h_2^2}{v^2}-\frac{1}{2}\Biggr].    
\end{equation}

\noindent With this we can now express the mass of $h_2$ as~\cite{PhysRevD.7.1888,Gildener:1976ih}

\begin{equation}
m_{h_2}^2 = \frac{d^2V_\text{eff}^\text{1-loop}}{d h_2^2}\Big|_v=8\beta v^2    
\end{equation}

\noindent where again we see for $\beta>0$ this is positive definite. Considering contributions from all standard and non-standard particles, we can write

\begin{equation}
m_{h_2}^2  = \frac{v_h^4}{8\pi^2 v^2}\Bigl(\lh^4+\frac{3}{8}g_{2}^4+\frac{3}{16}g_{2}^4\left(g_2^2+g_1^2\right)^2+3 g_X^4\left(v_s/v_h\right)^4-3y_t^4\Bigr)
%\frac{1}{8\pi^2 v^2}\Bigl(m_{h_1}^4+6m_W^4+3m_Z^4+3m_X^4-12m_t^4\Bigr)
\label{eq:mh2}
\end{equation}

\noindent where $g_2$ and $g_1$ are the gauge coupling corresponding to the SM groups $SU(2)_L$ and $U(1)_Y$ gauge respectively, and $y_t$ is the SM top Yukawa coupling\footnote{Recent measurement~\cite{CMS:2020djy} yields a best fit value of $y_t=1.16^{+0.24}_{-0.35}$. Unless otherwise mentioned, we have considered $y_t=1$.}. 
%where $m_{W,Z,t}$ are the masses of SM weak gauge bosons and top quark respectively. 
Also note, the fermion contribution appears with a relative sign. From Eq.~\eqref{eq:mh2} it is very important to note the role of the $U(1)_X$ gauge boson, without which $m_{h_2}^2<0$ (or equivalently $\beta<0$). Before concluding this section, let us summarize the present model. Aside from the massive dark gauge boson, the scalar particle spectrum consists of the CP-even state $h_1$, and an additional CP-even scalar $h_2$, which is a pNGB of scale symmetry breaking, with radiatively induced mass.

%%%%%%%%%%%%%%%%%%%%%%%%%
\subsection{Perturbative constraints \& RGE}\label{sec:rge}
%%%%%%%%%%%%%%%%%%%%%%%%%

There are three extra renormalizable couplings in this model on top of the SM Higgs quartic coupling $\lambda_H$

\begin{equation}
\{\lambda_S,\,\lambda_{HS},\,g_X\}.    
\end{equation}

The 1-loop $\beta$-functions extracted using {\tt PyR@TE}~\cite{Sartore:2020gou}, are given by

\begin{figure}[H]
    \centering
    \includegraphics[scale=0.47]{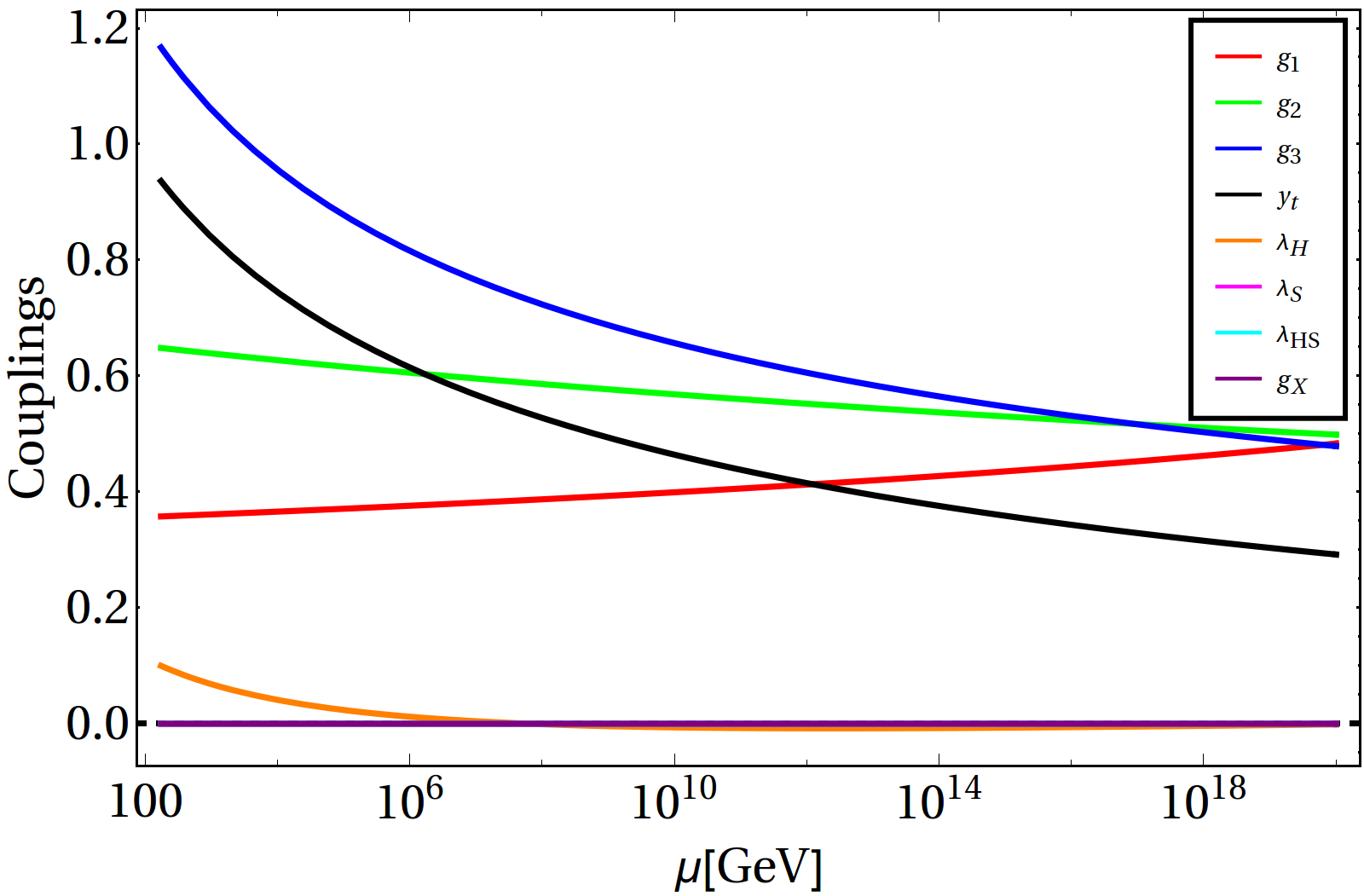}
    \caption{ \it RG evolution of different couplings appearing in the theory. The Higgs self-coupling turns negative at $\mu\gtrsim 10^9$ GeV, signalling that the SM Higgs potential becomes unstable below the Planck scale. We choose $m_t=173.2$ GeV.}
    \label{fig:run}
\end{figure}

\begin{equation}\begin{aligned}
&\left(4\pi\right)^2\,\beta\left(\lambda_H\right)=24 \lambda_H^{2}
+ \lambda_{HS}^{2}- 3 g_1^2 \lambda_H- 9 g_2^{2} \lambda_H+ \frac{3}{8} g_1^{4}+ \frac{3}{4} g_1^2 g_2^{2}+ \frac{9}{8} g_2^{4}+12\lambda_H y_t^2-6y_t^4
\\&
\left(4\pi\right)^2\,\beta\left(\lambda_S\right)=20 \lambda_S^{2}+ 2 \lambda_{HS}^{2}- 12 g_X^{2} \lambda_S+ 6 g_X^{4}
\\&
\left(4\pi\right)^2\,\beta\left(\lambda_{HS}\right)=12 \lambda_H \lambda_{HS}+ 8 \lambda_{HS} \lambda_S- 4 \lambda_{HS}^{2}-  \frac{3}{2} g_1^2 \lambda_{HS}- 6 g_X^{2} \lambda_{HS}- \frac{9}{2} g_2^{2} \lambda_{HS}+6y_t^2
\\&
\left(4\pi\right)^2\,\beta\left(g_X\right)=\frac{1}{3} g_X^{3} 
\end{aligned}\label{eq:beta1}
\end{equation}

\noindent where we define $\beta\left(X\right)\equiv\mu\,\frac{dX}{d\mu}$. The renormalization group (RG) equation for the top Yukawa coupling $y_t$ is 

\begin{equation}
\left(4\pi\right)^2\,\beta\left(y_t\right)=-8g_3^2y_t-\frac{9}{4}g_2^2y_t-\frac{17}{12}g_1^2y_t+\frac{9}{2}y_t^3.
\label{eq:beta2}
\end{equation}

\noindent To solve the RG equations and determine the RG evolution of the couplings of our models, we specify the initial conditions for the SM coupling constants at $m_t$ (top Yukawa coupling $y_t$) and the SM gauge couplings initial values are taken from~\cite{Buttazzo:2013uya}. Having thus specified the initial conditions for all couplings at the low scale $\mu=m_t$, we run them up to the high scale $\mu=\mpl$. As we see from Eq.~\eqref{eq:beta1}, $\beta\left(\lambda_S\right)\sim +g_X^4$, $\beta\left(\lambda_{HS}\right)\sim -g_X^2$ and $\beta\left(g_X\right)\sim g_X^3$, hence for a very small $g_X$, which is required for freeze-in production of the DM, the dark gauge coupling almost remains fixed over the scale $\mu$, which also influences the running of the coupling $\lhs$. Thus, although $\lhs$ contributes positively to the beta function of $\lh$, but that does not help to improve the Higgs vacuum.

%%%%%%%%%%%%%%%%%%%%%%%%%%%
%\subsection{Dynamical Generation of Weak Scale}
%%%%%%%%%%%%%%%%%%%%%%%%%%

%%%%%%%%%%%%%%%%%%%%%%%%%%%
\section{Freeze-in production of the Dark Matter}
\label{sec:dm-pheno}
%%%%%%%%%%%%%%%%%%%%%%%%%%%%

The $\mathbb{Z}_2$-odd gauge boson of the dark $U(1)_X$ gauge group is the DM candidate for the present model. In this section we aim to explore the viable parameter space where the DM can be produced out of equilibrium from the SM bath via freeze-in mechanism. As we have established in the model framework, the minimum mass of the $U(1)_X$ gauge boson, as decided by the scale invariance of the theory, is no less than 240 GeV. This clearly prevents the DM production from decay of the scalar states and only 2-to-2 annihilation of the bath particles are feasible way to account for the DM production. 

\begin{figure}[htb!]
    \centering
    \includegraphics[scale=0.35]{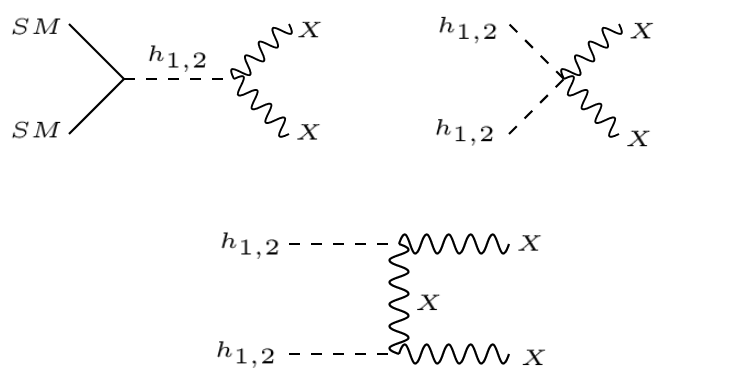}
    \caption{ \it Feynman graphs for DM yield from 2-to-2 annihilation of the SM particles. Here `SM' stands for all the SM fermions, scalar and gauge bosons.}
    \label{fig:feyn}
\end{figure}

\noindent The Feynman diagrams that contribute to the freeze-in production of the DM from the scattering of bath particles are shown in Fig.~\ref{fig:feyn}. We implemented this model in {\tt LanHEP}~\cite{Semenov:2008jy} and utilized the model files to calculate cross-sections via {\tt CalcHEP}~\cite{Belyaev:2012qa}. In the following sections we will determine the DM yield and subsequently the present DM abundance by numerically solving the Boltzmann equation, computing contributions due to these processes. 

%\begin{figure}
%\centering
%\includegraphics[scale=0.45]{figure/mXcplng.png}
%\caption{ \it $m_s/m_X$ as function of $\alpha_X(g_X)$ following Eq.~\eqref{eq:mx-gx}.}
%\label{fig:mxgx}
%\end{figure}

%%%%%%%%%%%%%%%%%%%%%%%%%%%%
\subsection{Freeze-in yield}\label{sec:freeze-in}
%%%%%%%%%%%%%%%%%%%%%%%%%%%%

As it is known, the key for freeze-in DM production is to assume that DM was not present in the early universe or the DM abundance was negligibly small after reheating. In case where the DM is produced via generic 2-to-2 annihilation, the Boltzmann equation (BEQ) reads~\cite{Duch:2017khv,Barman:2020ifq}

\begin{equation}
\begin{split}
x H s \frac{dY_X^\text{ann}}{dx} &= \gamma_\text{ann},
\end{split}\label{eq:beq}
\end{equation}

%\begin{equation}
%\begin{split}
%x H s \frac{dY_X^\text{D}}{dx} &= \gamma_D,
%\end{split}\label{eq:beq-dec}
%\end{equation}

where

\begin{equation}\begin{aligned}
\gamma\left(a,b\to1,2\right)&=\int\prod_{i=1}^4 d\Pi_i \left(2\pi\right)^4 \delta^{(4)}\biggl(p_a+p_b-p_1-p_2\biggr)f_a{^\text{eq}}f_b{^\text{eq}}\left|\mathcal{M}_{a,b\to1,2}\right|^2\\&=\frac{T}{32\pi^4}g_a g_b \int_{s_{min}}^\infty ds~\frac{\biggl[\bigl(s-m_a^2-m_b^2\bigr)^2-4m_a^2 m_b^2\biggr]}{\sqrt{s}}\sigma\left(s\right)_{a,b\to1,2}K_1\left(\frac{\sqrt{s}}{T}\right)\label{eq:gam-ann}, \end{aligned}    \end{equation}

\noindent with $a,b(1,2)$ as the incoming (outgoing) states and $g_{a,b}$ are corresponding degrees of freedom. Here $f_i{^\text{eq}}\approx\exp^{-E_i/T}$ is the Maxwell-Boltzmann distribution. The Lorentz invariant 2-body phase space is denoted by: $d\Pi_i=\frac{d^3p_i}{\left(2\pi\right)^3 2E_i}$. The amplitude squared (summed over final and averaged over initial states) is denoted by $\left|\mathcal{M}_{a,b\to1,2}\right|^2$ for a particular $2\to2$ scattering process. The lower limit of the integration over $s$ is $s_{min}=\text{max}\biggl[\left(m_a+m_b\right)^2,\left(m_{h_1}+m_2\right)^2\biggr]$. Here we define the DM yield $Y_X=n_X/s$ as the ratio of DM number density to the comoving entropy density in the visible sector since the DM is only produced from the SM bath. The parameter $x=m_X/T$ describes the SM sector temperature $T$, the Hubble parameter is denoted by $H$ and $\gamma=\langle\sigma v\rangle n_\text{eq}^2$ is the so-called reaction density~\cite{Chu:2011be} for the SM particles annihilating into the DM. All the relevant 2-to-2 annihilation cross-sections as a function of $s$ are collected in Appendix~\ref{sec:app-ann}, and as we see, all of them are proportional to $g_X^4$ in the lowest order of $g_X$. As mentioned before, the dark and visible sector communicates only via the portal coupling $\lambda_{HS}$, which in the scale invariant framework is not a free parameter (Eq.~\eqref{eq:cplngs2}). Thus, the dependence of the annihilation cross-section on the dark gauge coupling $g_X$ arises through the portal coupling. 

\begin{figure}[htb!]
$$    \includegraphics[scale=0.35]{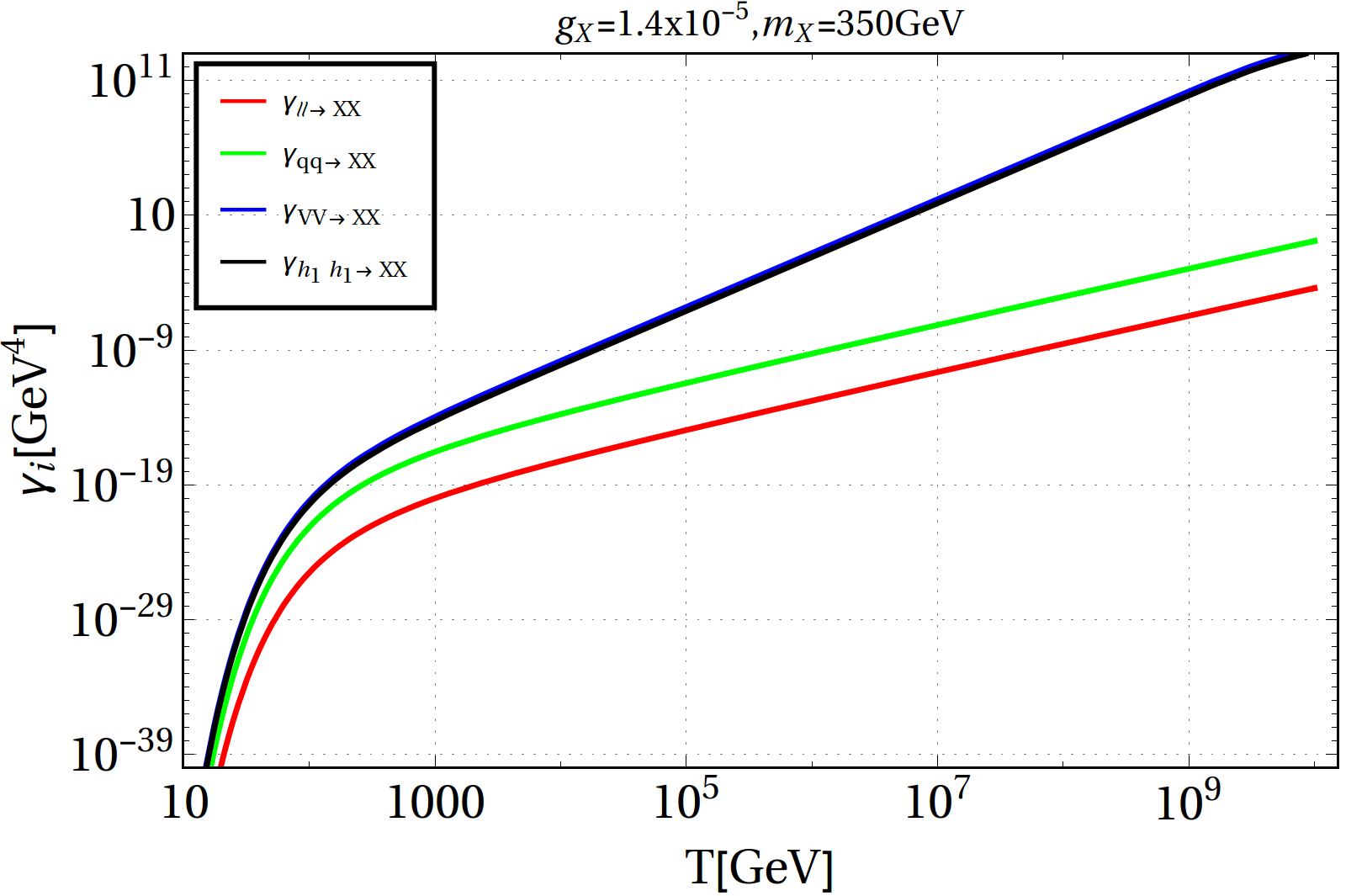}~~
    \includegraphics[scale=0.35]{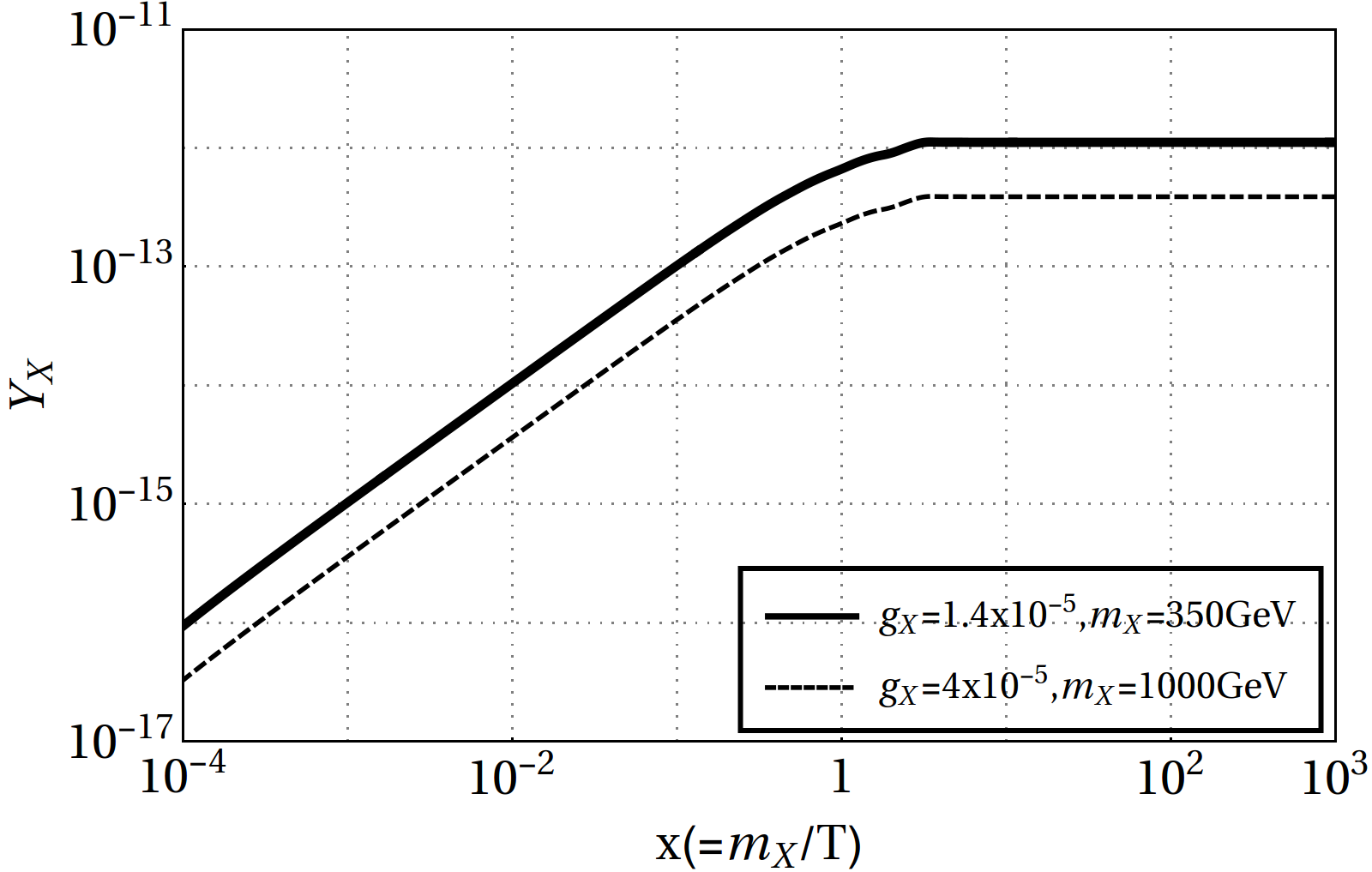}
    $$
    $$
   \includegraphics[scale=0.35]{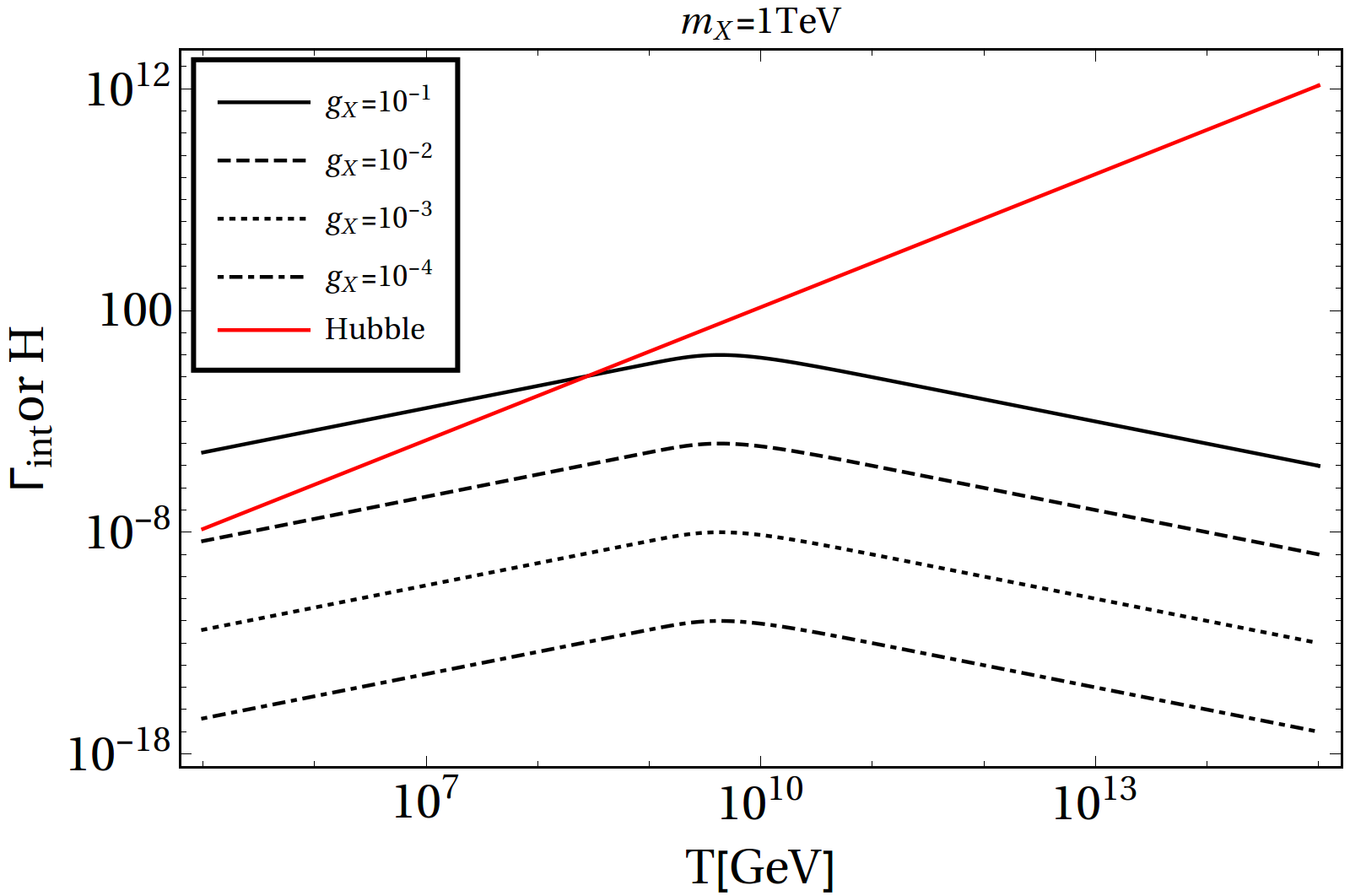}~~
    \includegraphics[scale=0.35]{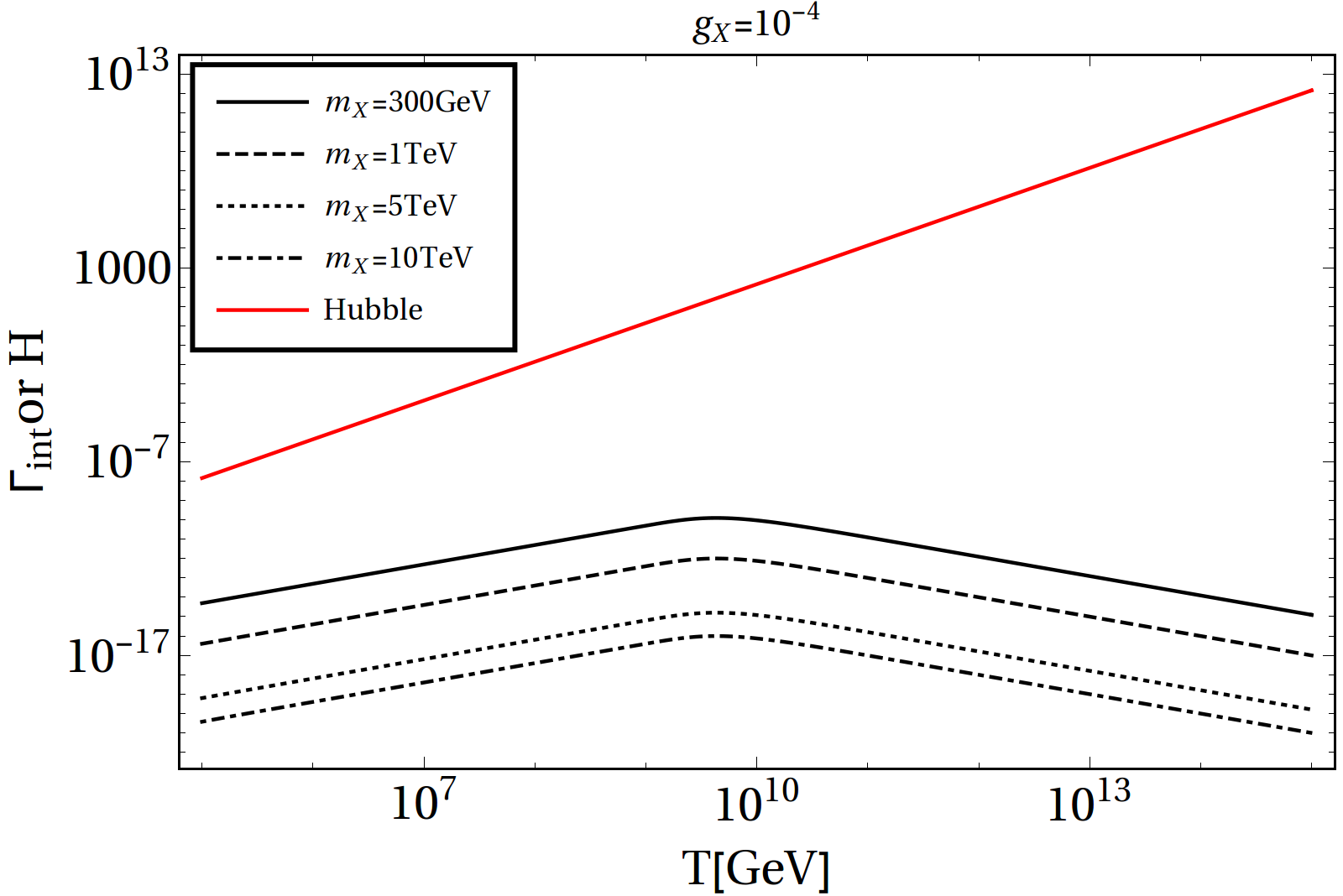}
    $$
    \caption{ \it  Top Left: Reaction densities as a function of the bath temperature, where different colours indicate different SM initial states. The parameters are chosen such that DM relic density is satisfied. Top Right: Evolution of DM yield with $x=m_X/T$ where the asymptotic yield produces observed relic abundance. Bottom Left: Comparison of DM freeze-in production rate with Hubble parameter for different choices of $g_X$ (chosen arbitrarily) with fixed $m_X=1$ TeV. Bottom Right: Same as bottom left but for a fixed $g_X=10^{-4}$ with different choices of DM mass.}
    \label{fig:yldplt}
\end{figure}

In the left panel of Fig.~\ref{fig:yldplt} we have shown the variation of the reaction densities as a function of the temperature $T$. The red and green curves correspond to SM leptons and quarks pair annihilating into the DM respectively, while the blue and black curves correspond to annihilation of SM gauge bosons and Higgs into a pair of DM. Note that, due to the presence of $SU(2)_L$ gauge coupling the contribution from the SM gauge bosons is maximum. On the other hand, the contact interaction (top right panel of Fig.~\ref{fig:feyn}) makes the scalar annihilation also important. Together, they imply, the contribution to DM relic abundance from gauge boson and scalar annihilations are dominant. In the right panel of Fig.~\ref{fig:yldplt} we illustrate the evolution of the DM yield as a function of the dimensionless quantity $x=m_X/T$. We choose two different benchmark points corresponding to two different DM masses such that the yield at $x\to\infty$ gives rise to the Planck~\cite{Planck:2018vyg} observed relic density. Here we see typical IR nature of the freeze-in, where the yield freezes in at $x\sim 1$ and then becomes constant up to the present temperature. Also note, for larger DM mass the yield is smaller which in turn demands a larger $g_X$ to obtain the observed abundance. This can be realized by considering an approximate analytical formulation of the IR yield for the DM~\cite{Hall:2009bx,Duch:2017khv,Barman:2020ifq}

\begin{equation}
Y_X \sim \sigma\left(T_\text{FI}\right)\, M_\text{pl}\,T_\text{FI}\sim \frac{g_X^4 M_\text{pl}}{m_X^5}\,m_{h_1}^4 
\label{eq:yx-approx}
\end{equation}

\noindent where the first relation comes out from simple dimensional arguments, and $\sigma\left(T_\text{FI}\right)$ is the production cross-section for the DM at freeze-in temperature $T_\text{FI}$ which in the present scenario can be approximated to be $\sigma\left(T_\text{FI}\right)\sim\lambda_{HS}^2/T_\text{FI}^2\sim\frac{g_X^4 m_{h_1}^4}{m_X^6}$. %$\sigma\left(T_\text{FI}\right)\sim g_X^4/T_\text{FI}^2$. 
We also consider $T_\text{FI}\sim m_X$ since for the SM plasma temperature below the threshold $m_X$, the yield ceases
to grow. From this approximate relation it is clear that in order to produce a fixed $Y_X$ which can provide the right abundance one has to choose a larger $g_X$ for a heavier DM. However, we emphasize that this is just an approximate relation, and we perform full numerical solution for solving the BEQ for yield. Before going into the details of the DM parameter space, few comments in order

%\ag{Can we add a bullet on why we say SI FIMP Miracle here ?}

\begin{itemize}
\item By demanding $m_{h_2}^2>0$ we put a lower bound on the vector DM mass: $m_X\gtrsim 250$ GeV for $g_X\lesssim 10^{-5}$, which is required for freeze-in production of the DM. This bound entirely arises from the scale-invariance of the theory. For such choices of parameters we find $m_{h_2}\ll m_X$. This ensures that it is not possible to produce the DM from the on-shell decay of the new scalar. The bound on $m_X$ also tells that the SM-like Higgs can not have on-shell decay into a pair of DM. Thus, the DM freeze-in production in the present framework takes place entirely via 2-to-2 scattering of the bath particles.

\item One must ensure the DM remains out of equilibrium in early times such that the freeze-in production condition $\Gamma_\text{int}<H$ is satisfied, where $\Gamma_\text{int}=\langle\sigma v\rangle_i n_\text{eq}^i$ is the reaction rate $(i\in\text{SM})$. We note, for $g_X\sim\mathcal{O}\left(10^{-4}\right)$ the reaction rate always falls below the Hubble rate $H$ ensuring the DM production takes place non-thermally at high  temperature. This is established from the bottom panel of Fig.~\ref{fig:yldplt}. Thus, for all benchmark values of $g_X$ and $m_X$ which give rise to right DM abundance, the DM is safely produced out of thermal equilibrium. 

\item As we will see, the observed DM abundance can be achieved for $g_X\sim\mathcal{O}\left(10^{-5}\right)$ for DM mass up to 2 TeV. For such $g_X$'s, we also find the scalar mixing $\sin\theta\sim\mathcal{O}\left(10^{-5}\right)$, which implies the portal coupling $\lambda_{HS}$ is extremely tiny, that in turn ensures that it is impossible for the dark sector to
equilibrate with the visible one. However, the non-thermalization of the dark sector by itself is not guaranteed. If the $XX\to h_2\,h_2$ process is very efficient then it will deplete the number density of $X$ and form a dark plasma with a temperature, which is in general, different from that of the SM temperature, within the dark sector. The non-thermalization condition within the dark sector can be encoded via

\begin{equation}
n_X\langle\sigma v_{XX\to h_2 h_2}\rangle\leq H    
\end{equation}

\noindent where $n_X$ is the number density of vector DM. Note that, the interaction cross-section $\langle\sigma v_{XX\to h_2 h_2}\rangle$ is proportional to the dark gauge coupling $g_X^4$, and thus the condition can be trivially satisfied for $g_X\sim\mathcal{O}\left(10^{-5}\right)$, which is of our interest.

\item The DM, in principle, can be produced in two different regimes: (a) symmetric phase: $T\gtrsim 160$ GeV, $\langle H\rangle=0$, where the SM gauge bosons are massless and  and (b) broken phase: $T<160$ GeV, $\langle H\rangle\neq0$ with massive SM gauge bosons. The total DM yield is then given by the sum of yields in these two regimes. Since the scale invariance of the theory induces symmetry breaking in both the sectors simultaneously, hence before the electroweak symmetry is broken, the DM is massless and {\it relativistic}, as opposed to the usual cases of freeze-in~\cite{Duch:2017khv,  Barman:2020ifq}. We aim to address this scenario in a future draft\footnote{A study of freeze-in production of scalar DM in the
relativistic regime can be found in~\cite{Lebedev:2019ton}.}.

\end{itemize}

\subsection{Viable dark matter parameter space}\label{sec:dm-paramspace}
%%%%%%%%%%%%%%%%%%%%%%%%%%%%

In order to compute the final DM yield, we have solved Eq.~\eqref{eq:beq} numerically, and obtain the relic abundance of the DM at the present epoch via

\begin{equation}
\Omega_X h^2 = \left(2.75\times 10^8\right) \left(\frac{m_X}{\text{GeV}}\right) Y_X(T_0)
\label{eq:relicX}    
\end{equation}

\noindent where $T_0$ is the temperature at the present epoch, which corresponds to $x\to\infty$. We must also remind that the  Planck~\cite{Planck:2018vyg} allowed relic density allows:

\begin{equation}
\Omega_\text{DM} h^2 = 0.11933\pm 0.00091,
\end{equation}

\noindent which we will use to constrain the relic density allowed parameter space. The presence of the (decaying) light scalar $h_2$ can give rise to two other very important bounds in the present model along with the relic density, which will constraint the model parameter space. We discuss them below. 

%In contrast to the usual freeze-in scenario, in the present context, spin-independent direct detection of the DM also plays a very crucial role in constraining the relic density allowed parameter space. Before going into the details of DM direct search we will first point out how big bang nucleosynthesis (BBN) can provide strong bound on the relic density allowed parameter space for the DM.

%%%%%%%%%%%%%%%%
\subsubsection{Bound from the Big Bang nucleosynthesis (BBN)}\label{sec:bbn}
%%%%%%%%%%%%%%%%

Within the DM freeze-in regime, the mediator $h_2$ is much lighter than the DM, and can be $\sim\mathcal{O}\left(1~\rm GeV\right)$ for DM mass $m_X\sim\mathcal{O}\left(10~\rm TeV\right)$. Also, its coupling to the SM fermions turns out to be $\theta\sim\mathcal{O}\left(10^{-5}\right)$. Since $h_2$ decay to SM particles is suppressed by the small mixing angle, its lifetime tends to be very long and it may cause problems in the early Universe~\cite{Zhang:2015era}. In particular there will be an upper bound on the $h_2$-lifetime from  nucleosynthesis.
%A light weakly coupled scalar with a thermal number density can decay appreciably during the BBN, and spoil the successful predictions of light element yields accumulated in the early universe. 
It is well known that quasi-stable particles with decay times $\tau\gtrsim 0.1$ seconds may significantly perturb the
primordial light element nucleosynthesis occurring approximately between 1 and 1000 seconds after the birth of the universe. The decay lifetime of $h_2$ into the SM states can thus potentially perturb the successful predictions of light element yields accumulated in the early universe. If the decay occurs after BBN with $\tau_{h_2}>1$ sec, entropy production has to be less than $\sim 10\%$
(assuming no change to light element abundances due to it) because the precision measurements
of the baryon density through BBN and CMB observations match very well~\cite{Steigman:2012ve,Kaplinghat:2013yxa}. Therefore, the absence of significant entropy production after BBN will put a strong bound on the lifetime $\tau_{h_2}$. If the mediator decays before BBN, entropy production constraints are
non-existent. 

A fully quantitative analysis of these effects is beyond the scope of this paper and can be found in, for example, Ref.\cite{Berger:2016vxi}. Instead we follow~\cite{Kawasaki:2000en} and, to remain within
the 2$\sigma$ limit of the observed $\textsuperscript{4}\text{He}$ abundance, require $\Gamma_{h_2}^{-1}\equiv\tau_{h_2}<1$ sec. Note that, this is rather a conservative bound given the fact that the light scalar never comes in equilibrium with the SM due to feeble portal coupling. For $m_{h_2}\sim\mathcal{O}\left(1~\rm GeV\right)$, $h_2$ can decay on-shell into a pair of light fermions, and also to photon and gluon final states via loop. The partial decay widths to the SM final states of $h_2$ are given by~\cite{Djouadi:2005gi,Krnjaic:2015mbs}

\begin{equation}\begin{aligned}
&\Gamma_{ff} = \frac{G_F\sin^2\theta\,N_c}{4\sqrt{2}}m_f^2 m_{h_2}\Biggl(1-\frac{4m_f^2}{m_{h_2}^2}\Biggr)^{3/2}
\\&
\Gamma_{\gamma\gamma}=\frac{G_F\sin^2\theta}{128\sqrt{2}}\frac{\alpha^2 m_{h_2}^3}{\pi^3}\Biggl|\sum_f N_c Q_f^2 \mathcal{A}_{1/2}\left(x_f\right)+\mathcal{A}_1\left(x_f\right)\Biggr|^2
\\&
\Gamma_{gg} = \frac{G_F\sin^2\theta}{36\sqrt{2}}\frac{\alpha_s m_{h_2}^3}{\pi^3}\Biggl|\frac{3}{4}\sum_q\mathcal{A}_{1/2}\left(x_q\right)\Biggr|^2
\end{aligned}\label{eq:decayW}
\end{equation}

\noindent where $x_i=m_{h_2}^2/4m_i^2$, $N_c$ is the number of colors for a given fermion species and $Q_f$ is its electromagnetic charge. Also,

\begin{equation}\begin{aligned}
&\mathcal{A}_{1/2}\left(x\right)=2\Bigl[x+\left(x-1\right)f(x)\Bigr]x^{-2}
\\&
\mathcal{A}_1(x)=-\Bigl[2x^2+3x+3\left(2x-1\right)f(x)\Bigr]x^{-2}
\end{aligned}
\end{equation}

with

\begin{equation}
f(x) \equiv
    \begin{cases}
        \text{Arc}\sin^2\sqrt{x} & x\leq 1,\\[8pt]
        -\frac{1}{4}\Bigl[\log\frac{1+\sqrt{1-x^{-1}}}{1-\sqrt{1-x^{-1}}}-i\pi\Bigr]^2 & x>1.
    \end{cases}
\end{equation}

%Since an in-depth BBN analysis is beyond the scope of this paper, we simply demand that the light scalar lifetime has to be $\Gamma_{h_2}^{-1}\equiv\tau\gtrsim\tau_\text{BBN}\simeq 1$ sec. 

Now, since $\Gamma_{h_2}$ depends both on the mixing $\sin\theta$ and on $m_{h_2}$, which is turn, are functions of $g_X$ and $m_X$, therefore it is evident that the requirement $\tau_{h_2}<1$ sec will constraint the resulting parameter space in the $g_X-m_X$ plane, especially ruling out small $g_X$ values which can result in very long lifetime of the light scalar. 

%%%%%%%%%%%%%%%%
\subsubsection{Dark matter direct detection}\label{sec:si-dd}
%%%%%%%%%%%%%%%%

%It is usually claimed that the DM direct-detection experiments do not provide relevant constraints for models in which the DM particles are mainly produced via the freeze-inmechanism as the DM nuclear recoil cross sections are suppressed by tiny portal couplings. 

Contrary to the usual notion that DM nuclear recoil cross sections are suppressed by tiny portal couplings (which is required for DM freeze-in), and therefore DM direct-detection experiments do not provide relevant constraints in freeze-in framework, in the present set-up the DM-nucleon scattering cross-section can be hugely amplified because it takes place via $t$-channel mediation of the light scalar $h_2$ (along with the SM-like Higgs $h_1$). The DM-nucleus scattering cross-section in this case has the form~\cite{Duch:2017khv}

\begin{equation}
\frac{d\sigma_{XN}}{dq^2}\left(q^2\right)=\frac{\sigma_{XN}}{4\mu_{XN}^2 v^2}\,F\left(q^2\right)
%\sigma_{XN} = \frac{\lambda_{HS}^2\,f_N^2\,m_X^2\,m_N^2\,\mu_{XN}^2}{\pi\,m_{h_1}^4\,m_{h_2}^2\,\left(m_{h_2}^2+4\mu_{XN}v^2\right)}
\label{eq:ddx}
\end{equation}

\noindent where $v_\text{DM}$ is the DM velocity in the lab frame and the DM-nucleus reduced mass is given by $\mu_{XN}=m_X\,m_N/\left(m_X+m_N\right)$, and

\begin{equation}
\sigma_{XN} =\frac{\lambda_{HS}^2\,f_N^2\,m_X^2\,m_N^2\,\mu_{XN}^2}{\pi\,m_{h_1}^4m_{h_2}^2\left(m_{h_2}^2+4\mu_{XN}^2\,v_\text{DM}^2\right)}    
\end{equation}

%{\bb{Note that we have written the direct search cross-section in terms of the portal coupling and not $g_X$. We can write this using the second relation in Eq.~\eqref{eq:cplngs}. We also use $m_2\ll m_{h_1}$.}}

\noindent is the total cross-section with the effective DM-nucleon coupling is $f_N\approx 0.3$~\cite{Cline:2013gha}. Here the factor $m_{h_2}^2+4\mu_{XN}^2 v_\text{DM}^2$ in the denominator represents the light mediator effects~\cite{Duch:2017khv,Duch:2019vjg}. The function $F(q^2)$ is defined as~\cite{Geng:2016uqt}

\begin{equation}
F(q^2) = \frac{\Bigl(1+q_\text{min}^2/m_{h_2}^2\Bigr)\Bigl(1+q_\text{ref}^2/m_{h_2}^2\Bigr)}{\Bigl(1+q^2/m_{h_2}^2\Bigr)^2}
%\frac{m_{h_2}^2}{\left(q^2+m_{h_2}^2\right)^2}\Bigl[m_{h_2}^2+4\mu_{XN}^2 v^2\Bigr]   
\end{equation}

\noindent which encodes the effects of the light mediator. Usually, $q_\text{min}^2$ is very small compared to the other scales appearing in the process, and thus can be taken to be zero, while $q_\text{ref}^2=4\mu_{XN}^2v_\text{DM}^2$ is related to the energy thresholds of DM direct detection experiments. In the limit $m_{h_2}^2\gg q^2\sim 4\mu_{XN}^2v_\text{DM}^2$, the form factor $F(q^2)\approx 1$ and we recover the conventional DM-nucleon scattering cross-section for contact interactions. But for  $m_{h_2}^2\ll q^2$, Eq.~\eqref{eq:ddx} will have extra $q^2$ dependence characterized by $F(q^2)$. We again remind the readers that the portal coupling $\lambda_{HS}$ is not a free parameter, rather can be expressed in term of the dark gauge coupling $g_X$ following the relations in Eq.~\eqref{eq:cplngs2}. This makes the DM-nucleus scattering cross-section $\sigma_{XN}\propto g_X^4$, which implies large $g_X$ will result in large $\sigma_{XN}$, making the parameter space vulnerable from direct detection bounds\footnote{In deriving the direct search bounds we do not consider RGE flow of $g_X$ from UV scale down to nuclear physics energy scale since it remains almost fixed to a very small value (set by freeze-in), see Fig.~\ref{fig:run}.}. We consider exclusion limits from XENON1T~\cite{XENON:2018voc} (black solid curve), and projected bounds from PandaX-4T~\cite{PandaX:2018wtu} (black dashed), LUX-ZEPLIN (LZ) (brown dashed)~\cite{LUX-ZEPLIN:2018poe}, XENONnT~\cite{XENON:2020kmp} (black dotdashed) and DARWIN~\cite{DARWIN:2016hyl} (black dotted) experiments which provide upper limit on the DM-nucleon scattering cross-section at 90\% C.L. 

\begin{figure}[htb!]
    \centering
    \includegraphics[scale=0.4]{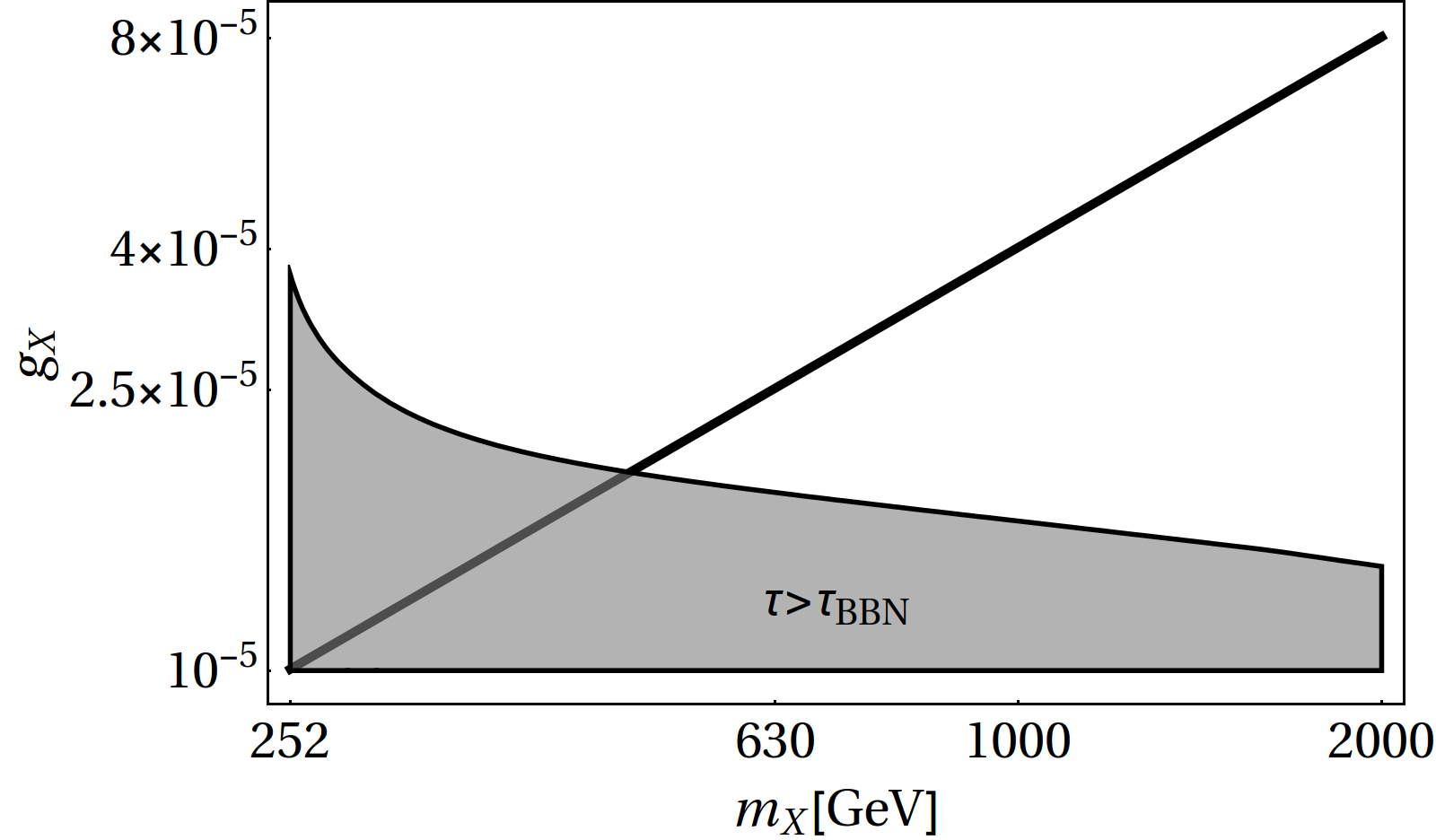}~~
    \includegraphics[scale=0.38]{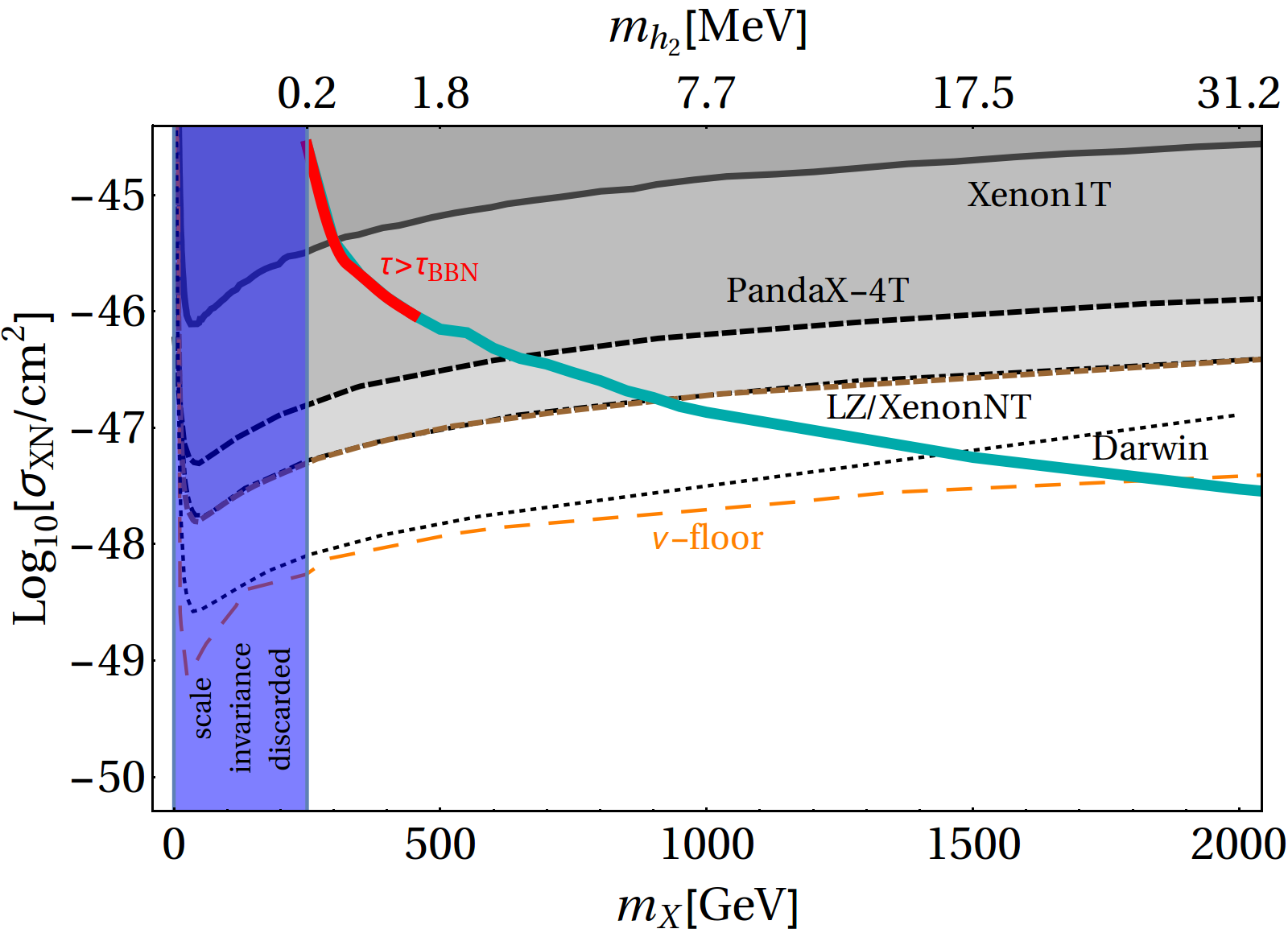}
    \caption{ \it Left: The black straight line represents the contour satisfying Planck observed relic density in the bi-dimensional plane of $g_X$-$m_X$. Right: Relic density allowed parameter space (cyan region) projected in direct search plane where exclusion limits from spin-independent direct search experiments are portrayed (black solid, black thick dashed, black dot-dashed, brown thick and black dotted curves) along with the expected discovery limit corresponding to the so called ``$\nu$-floor" from CE$\nu$NS of solar and atmospheric neutrinos for a Ge target (orange dashed curve).}
    \label{fig:paramsp}
\end{figure}

We can now proceed to explore the allowed parameter space for the DM satisfying relic density and direct search limits. This is furnished in Fig.~\ref{fig:paramsp}. The black thick straight line in the left panel of Fig.~\ref{fig:paramsp} shows the contour satisfying observed relic abundance. Here we see, with increase in DM mass, one needs a larger $g_X$ to satisfy the relic constraint $\Omega_X h^2\simeq 0.12$. This can again be explained from the approximate relation in Eq.~\eqref{eq:yx-approx}, from which we find $\Omega_X h^2\propto m_X\times Y_X \propto \left(g_X/m_X\right)^4$. Thus, it is evident, one needs to tame down the gauge coupling in order to satisfy the observed abundance for a lighter DM, and the dependence is linear. %{\bb{small bulge around 600 GeV is numerical glitch}}.
We see, up to a DM mass of $m_X\sim 2$ TeV, right abundance is obtained in the vicinity of $g_X\sim 10^{-5}$. A part of the relic density allowed parameter space is also ruled out from BBN constraint. This is shown by the gray region. As it is seen, for a fixed $m_X$, increasing $g_X$ relaxes the BBN bound since it gives rise to a larger mixing (and also a larger $m_{h_2}$ following Eq.~\eqref{eq:mh2}) that results  in a smaller lifetime (Eq.~\eqref{eq:decayW}). The same effect is also realized on increasing $m_X$ for a fixed $g_X$.  The BBN bound thus discards DM mass below $m_X\lesssim 467$ GeV with $g_X\lesssim 2\times 10^{-5}$. In other words, a lower bound on $g_X$ can be set solely from BBN, depending on the corresponding DM mass. 

%Thus, this typical pattern of the gray region i.e., where $\tau_{h_2}> 1$ sec, is a result of the scale invariant nature of the theory.

%BBN bound on the lifetime of $h_2$ allows $g_X\gtrsim 2\times 10^{-5}$ which corresponds to a DM mass $m_X\gtrsim 467$ GeV. 

In the right panel we show how much of the relic density allowed parameter space survives spin-independent direct search constraints for DM mass up to 2 TeV.
%Here we see, with the decrease in $g_X$ the DM-nucleon scattering cross-section decreases following Eq.~\eqref{eq:ddx}. Thus, the red, green and blue curves easily evade exclusion limits from not only PandaX-4T but XENONnT as well. However, for such tiny $g_X$ the DM is hugely under abundant within the given range of DM mass. 
The right DM abundance is obtained within the cyan band where we have $g_X\in\{1\times 10^{-5}-8\times 10^{-5}\}$, as we can see from the left panel. The BBN bound can also be projected in $\sigma_{XN}-m_X$ plane, as shown by the red points. Since with the increase in $g_X$ the DM-nucleon scattering cross-section increases following Eq.~\eqref{eq:ddx}, hence one expects comparatively lower DM mass region to be safe from direct detection bound where one requires lower $g_X$ to obtain the right abundance. However, a lower $g_X$ also results in a lighter $m_{h_2}$ that enhances the direct search cross-section in turn. Hence, we see, lower DM masses are tightly constrained from several direct search experiments, while the bounds get relaxed as we move on to heavier DM mass. Within this range of $g_X$ and $m_X$ that produces right abundance, the mass of the new scalar turns out to be $m_{h_2}\sim 30$ MeV (mentioned along the top horizontal axis), which is of the order of typical momentum transfer scale $q\sim 10$ MeV in DM-nucleus scattering. We see, the viable region of the parameter space lies below the PandaX-4T exclusion limit for DM mass $m_X\gtrsim 642$ GeV with $g_X\gtrsim 2.7\times 10^{-5}$. For $g_X\gtrsim 1.2\times 10^{-5}$ (corresponding $m_X\gtrsim 300$ GeV) the relic density allowed parameter space lies below the exclusion limit from XENON1T, however, a major part of the parameter space is ruled out by BBN (shown in red). One should note here, it is possible to obtain the right relic abundance for a DM of mass as large as $\sim\mathcal{O}(10~\rm TeV)$ with $g_X\sim\mathcal{O}\left(10^{-4}\right)$, which, needless to mention, will satisfy the direct search bound. In that case $m_{h_2}$ can be as large as $\sim\mathcal{O}(1~\rm GeV)$. Hence, it is possible to have the scalar mass around the GeV scale in the expense of making the DM heavy, satisfying both relic abundance and direct detection constraints. As we shall see in the next section, such light scalars can be probed in various ongoing and proposed experiments. A large part of the parameter space, on the other hand, can be potentially tested in the future XENONnT experiment. We also show limit from the so called ``$\nu$-floor"~\cite{Billard:2013qya} via the orange dashed lines, below which the number of neutrino events due to coherent elastic neutrino-nucleus scattering (CE$\nu$NS) is expected to be much larger than the number of DM events, which prevents to identify DM signals with certainty. The parameter space of our interest lies just below the $\nu$-floor for DM mass $m_X\gtrsim 1.8$ TeV. 

%In Ref.~\cite{Duch:2017khv} the authors have discussed in details about several indirect search possibilities of a freeze-in abelian vector DM. These signatures typically depend on the decay of $h_2$ into electron-positron pair and into gamma ray final states.

Before closing this section we would like to mention that in the present model the $t$-channel scalar mediation can as well give rise to non-zero DM-electron scattering cross-section. As it is known, experimental sensitivity to events with a single or a few ionization electrons has been demonstrated with XENON10\cite{Essig:2017kqs} and more recently DarkSide~\cite{DarkSide:2018ppu}, thus allowing for a novel way to probe light DM scattering off of electrons with an existing experimental setup~\cite{Essig:2012yx}. Now, an approximate analytical expression for DM-$e$ scattering cross-section has the form: $\sigma_{Xe}\propto\lhs^2\,g_X^2\,\Bigl(\frac{m_X^2\,m_e}{m_{h_1}^2\,m_{h_2}^2}\Bigr)^2$, which produces a cross-section $\sim\mathcal{O}\left(10^{-50}\right)\text{cm}^2$ for a DM mass of 300 GeV with $g_X=10^{-5}$ (where $\lhs$ can be substituted using Eq.~\eqref{eq:cplngs2}). Such a small cross-section is far below the present sensitivity of DM-electron scattering experiments, hence we do not consider them here.

%{\bb{For electron scattering experiments, the DM-$e$ scattering cross-section is roughly proportional to $\left(\lambda_{HS}\, m_e\right)^2$, which is extremely tiny due to the presence of both small portal coupling $\lambda_{HS}$ and $m_e$.}}

%%%%%%%%%%%%%%%%%%%%%%%%%%%%%%%
\section{Laboratory Phenomenology}
\label{sec:expts}\vspace{0.5cm}
%%%%%%%%%%%%%%%%%%%%%%%%%%%%%%%

\begin{figure}[H]
%    \centering
    $$
    \includegraphics[scale=0.5]{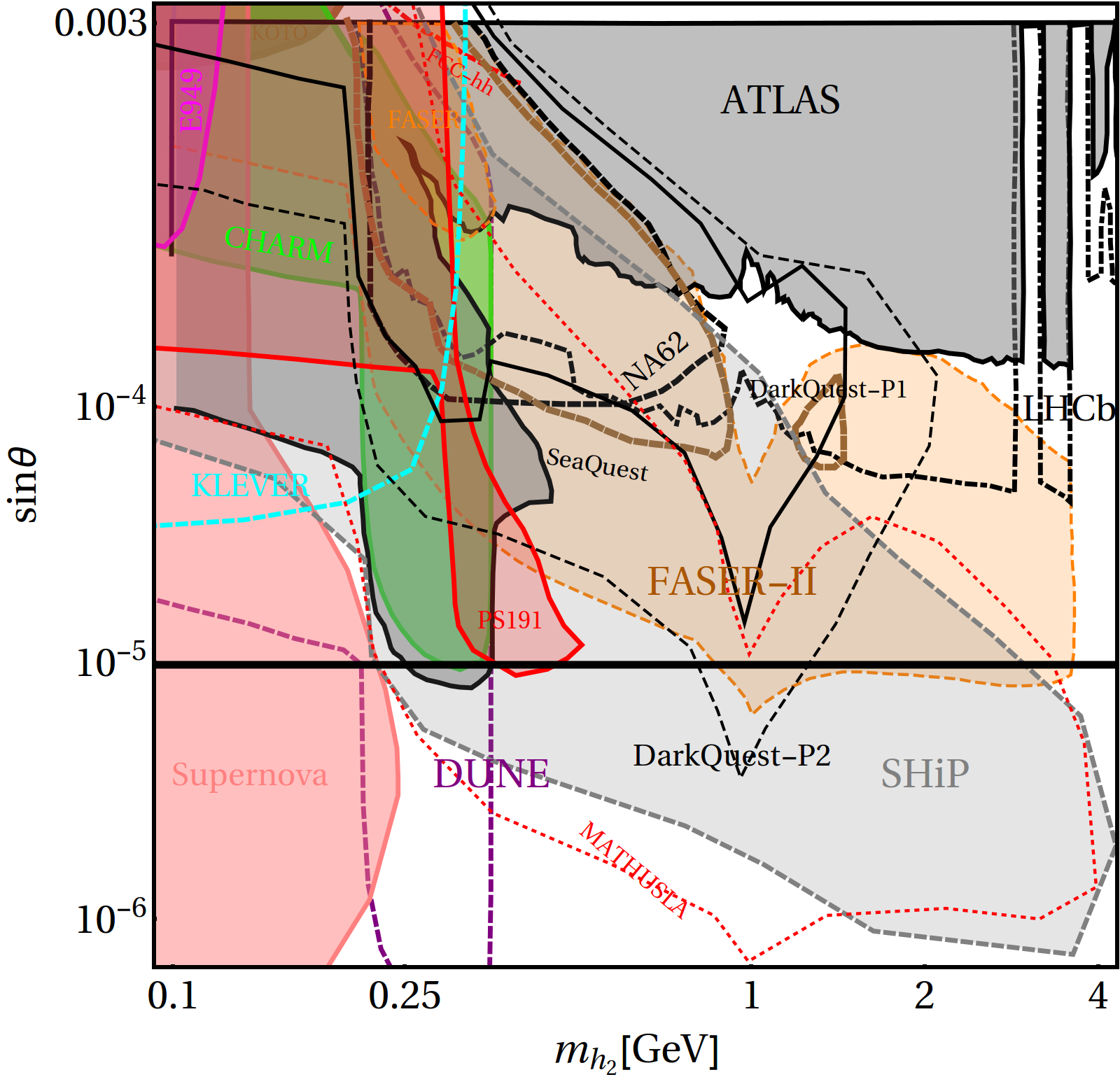}
    $$
    % $$
    % \includegraphics[scale=0.53]{figure/barPlt.png}
    % $$
    \caption{ \it The thick black straight line shows the DM parameter space complying with the Planck observed relic abundance and satisfying spin-independent direct search exclusion limit in $\sin\theta-m_{h_2}$ plane. Experimental limits are shown from E949~\cite{E949:2008btt}, CHARM~\cite{CHARM:1985anb}, NA62~\cite{Gonnella_2017}, FASER-I\&II~\cite{Feng:2017vli, FASER:2018eoc, FASER:2018bac, FASER:2019aik}, FCC-hh~\cite{FCC:2018vvp, DEV2017179}, ATLAS~\cite{ATLAS:2012qaq,ATLAS:2014jdv, Chalons:2016jeu, Robens:2015gla}, SeaQuest~\cite{Berlin:2018pwi}, LHCb~\cite{Gligorov:2017nwh}, KLEVER~\cite{Moulson:2018mlx}, DUNE~\cite{DUNE:2015lol, Berryman:2019dme}, DarkQuest-Phase2~\cite{Batell:2020vqn}, MATHUSLA~\cite{Curtin:2018mvb}, SHiP~\cite{SHiP:2015vad} and PS191~\cite{Bernardi:1985ny, Gorbunov:2021ccu}. All proposed experimental limits are denoted by dashed curves, while the existing limits are in solid. Exclusion region from supernova cooling and BBN are also depicted}
    \label{fig:exptLim1}
\end{figure}

To this end, we showed that the freeze-in vector DM in our model can give rise to a viable parameter space satisfying bounds from relic density. We have also realized that the scale invariance of the theory gives rise to a naturally light scalar mediator which influences the nuclear recoil spectrum in direct detection experiment, leading to the possibility of probing this model there-in. In this section we investigate the prospects of searching for this light scalar (and thus probing the DM parameter space) in intensity frontier and lifetime frontier experiments which look for for light, weakly
interacting, electrically neutral long-lived particles. These experiments are capable of probing extremely small mixing angles. Dark sectors with
light degrees of freedom can be probed with a variety of
experiments at the luminosity frontier, including proton~\cite{Batell:2009di, deNiverville:2011it, deNiverville:2012ij, Kahn:2014sra, LBNE:2013dhi, Soper:2014ska, Dobrescu:2014ita, Coloma:2015pih, deNiverville:2016rqh, MiniBooNE:2017nqe, MiniBooNEDM:2018cxm, MATHUSLA:2018bqv, FASER:2018bac}, electron~\cite{Bjorken:2009mm, Izaguirre:2013uxa, Diamond:2013oda, Izaguirre:2014dua, Batell:2014mga, BaBar:2017tiz, Berlin:2018bsc, Banerjee:2019pds} and positron fixed target facilities~\cite{Accardi:2020swt}. Here we show, the allowed parameter space of the DM lies well within the reach of several such experiments given the mass and mixing of the light scalar\footnote{Light dark sector scenarios have been explored in great detail over the past decade~\cite{Bramante:2016yju, Alexander:2016aln, Battaglieri:2017aum, Darme:2017glc, Winkler:2018qyg, Egana-Ugrinovic:2019wzj, Okada:2019opp, Foroughi-Abari:2020gju, Agrawal:2021dbo,Nardi:2018cxi}.}. \textit{The noteworthy point here is that, the coupling responsible for the freeze-in production is being directly probed at the experimental level, which, in general, is not the case\footnote{For probing freeze-in in collider experiments the presence of a lightest odd sector particle (LOSP) is generally required~\cite{Hall:2009bx}.}.} However, given the underlying scale invariance, the resulting parameter space is extremely constrained and predictive. 

\begin{figure}[H]
$$
\includegraphics[scale=0.55]{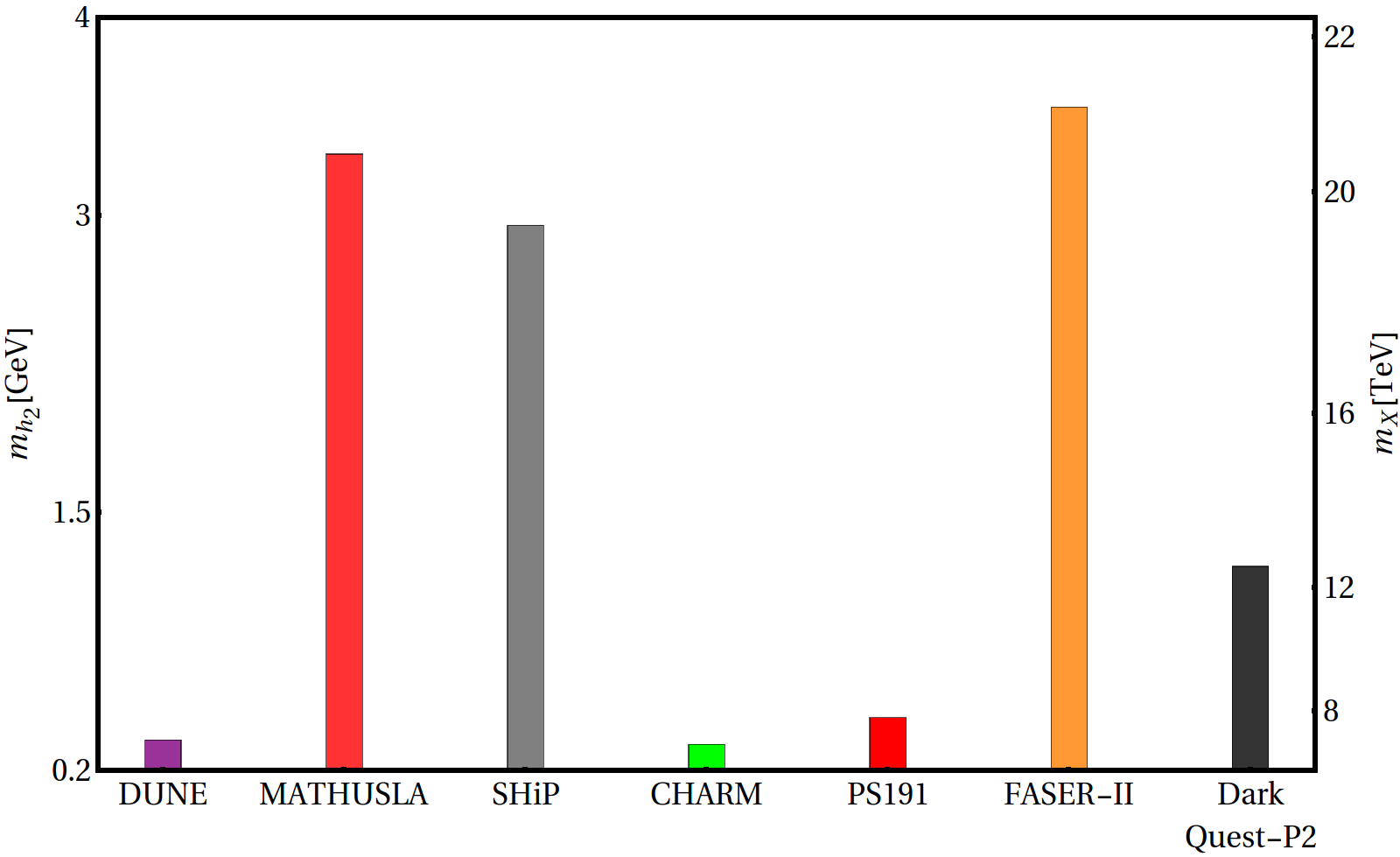}
$$
    \caption{ \it Different colored bars represent the reach of the corresponding experiments (mentioned along the horizontal axis) in probing the maximum mass of the light scalar. The DM mass, satisfying relic density and spin-independent direct detection bound, corresponding to different $m_{h_2}$ (following Eq.~\eqref{eq:mh2}) are mentioned along the vertical axis on the right side.}
    \label{fig:exptLim2}
\end{figure}

Let us now note that the requirement of $\Omega_X h^2\simeq 0.12$ typically constraints the ratio $m_X/g_X$ since $\Omega_X h^2\propto\left(g_X/m_X\right)^4$, as mentioned before (c.f. Eq.~\eqref{eq:yx-approx}). This also implies, for fixed a  $\Omega_X h^2$, the mixing becomes constant since $\sin\theta= v_h\Big/\sqrt{v_h^2+\left(m_X/g_X\right)^2}$ and $g_X/m_X$ is determined from relic abundance. Thus, the scale-invariance of the theory, together with the requirement of right relic abundance, fixes $\sin\theta$ to a constant value. For all choices of $\{m_X,g_X\}$ that leaves the combination $m_X/g_X$ constant (proportional to $\Omega_X h^2$), we find $\sin\theta\sim 10^{-5}$. This is what we are referring to as a ``miracle" in the present model, as the value of this mixing angle is always fixed, as decided by the scale invariance of the theory. Thus, in the $m_{h_2}-\sin\theta$ plane, the relic density allowed parameter space is simply a straight line with a fixed $\sin\theta$ (independent of $m_{h_2}$), determined from the observed relic density. This is shown by the black horizontal straight line in the top panel of Fig.~\ref{fig:exptLim1}. For the given mass range of $m_{h_2}$ the constraint from spin-independent direct search is trivially satisfied since $m_{h_2}\geq 0.1$ GeV corresponds to DM mass $m_X\gtrsim$ 3.5 TeV. 

\begin{figure}[H]
$$
\includegraphics[scale=0.45]{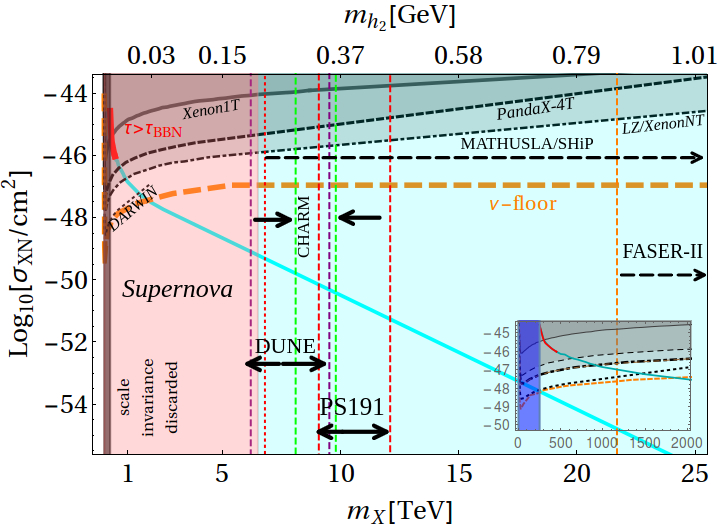}
$$
    \caption{ \it Summary of all experimental bounds described in Sec.~\ref{sec:dm-paramspace} and Sec.~\ref{sec:expts} in the direct search plane, where the relic density allowed DM parameter space shown via the thick cyan curve as in Fig.~\ref{fig:paramsp}. The pink region is disallowed from supernova bound on the mass of light scalar. The light blue region in the background denotes the mass range that can only be probed in lifetime frontier experiments. The colorful vertical dashed lines indicate the ranges in which different intensity frontier experiments can search. Inset shows the sensitivity of direct detection experiments up to DM mass of 2 TeV (Fig.~\ref{fig:paramsp}).}
    \label{fig:summaryPlt}
\end{figure}

%For DM as massive as 20 TeV, corresponding $g_X=7.6\times 10^{-4}$ to match the observed abundance. 
%\ag{We need to mention somewhere that these experiments are basically "singlet scalar searches". Which we have conveniently converted to light scakar hunt for our DM model purposes.}

In Fig.~\ref{fig:exptLim1} we summarize the result obtained from DM phenomenology, together with the search reach of several planned/proposed experiments (contours with the names of experiments indicated) in $m_{h_2}-\sin\theta$ plane. We see, light scalar mediator in the present model is within the reach of CHARM~\cite{CHARM:1985anb}, DUNE~\cite{DUNE:2015lol, Berryman:2019dme},  FASER-I\&II~\cite{Feng:2017vli, FASER:2018eoc, FASER:2018bac, FASER:2019aik}, PS191~\cite{Bernardi:1985ny, Gorbunov:2021ccu}, DarkQuest-Phase2~\cite{Batell:2020vqn}, MATHUSLA~\cite{Curtin:2018mvb} and SHiP~\cite{SHiP:2015vad} (see for example~\cite{Beacham:2019nyx} for a summary on these experiments) for the allowed range of mass and mixing. Note that, the weakly coupled light scalar mediator can be produced on shell during a supernova (SN) explosion and significantly contribute to its energy loss, thereby shortening the duration of the observable neutrino pulse emitted during core collapse. The most significant such constraint arises from SN1987A~\cite{Krnjaic:2015mbs, Dev:2020eam}, that excludes $m_{h_2}\sim 148$ MeV, as shown by the pink region labelled as `Supernova.'  Fig.~\ref{fig:exptLim2} shows the sensitivity of the potential experiments (as obtained from the left panel) in probing different ranges of $m_{h_2}$. For each $m_{h_2}$ it is possible to choose suitable $\{m_X,g_X\}$ (mentioned along the right vertical axis) such that right DM relic abundance is obtained abiding all bounds discussed previously.   

%{\bb{write description}}

In Fig.~\ref{fig:summaryPlt}, we have summarized the bounds from both DM direct search and lifetime frontier experiments that we have discussed in Sec.~\ref{sec:dm-paramspace} and Sec.~\ref{sec:expts} respectively. The thick cyan curve denotes the relic density allowed parameter space for the DM as we already found in Fig.~\ref{fig:paramsp}. We have now extended this up to $\sim 25$ TeV to accommodate the bounds on $m_{h_2}$ from lifetime frontier experiments. We show the exclusion limits from spin-independent direct detection experiments from XENON1T, and projected sensitivity limits from PandaX-4T, LUX-ZEPLIN (LZ), XenonNT and DARWIN. As expected, for low DM mass region (below 1 TeV) these bounds are severe but become rather weak for heavier DM mass\footnote{Direct search limit on superheavy DM have been explored in Refs.~\cite{Kavanagh:2017cru, Bramante:2018qbc, Clark:2020mna}.}. A light scalar with mass below $\sim 250$ MeV is ruled out from supernova observations as indicated by the pink region. This corresponds to a DM mass $\lesssim 6.4$ TeV. Above DM mass of $\sim 1.8$ TeV the relic density allowed parameter space gets submerged into the $\nu$-floor, where separating DM scattering from the background neutrino scattering is rather challenging. However, it is interesting to note that in those regions several intensity frontier experiments can provide excellent sensitivity. The vertical colored dashed lines show the range of $h_2$ masses (equivalently DM mass) in experiments like DUNE, CHARM, PS191, MATHUSLA, SHiP, FASER-II etc. Thus, even though direct search experiments may lose sensitivity in exploring heavy DM masses, but the presence of the light scalar can still provide potential signals in the plethora of intensity and lifetime frontier experiments, and thanks to the scale invariance of the theory, this automatically implies a signature of the freeze-in DM, in this model. 

%%%%%%%%%%%%%%%%%%%
%\begin{comment}
%%%%%%%%%%%%%%%%%%%%%%%%
\section{Discussion and Conclusion}
\label{sec:concl}
%%%%%%%%%%%%%%%%%%%%%%%

Scale invariance at the classical level leads to alleviating the gauge hierarchy problem as the observed scales (EW and DM scales) can be dynamically generated at the quantum level, removing the perilous quadratic divergences from the UV cut-off. In a conformal gauge extension of the SM we showed that the observed cosmological DM abundance can also be satisfied. We investigated the freeze-in production of dark matter (DM), which rests on the possibility of DM particle being only feebly interacting with the visible SM sector; here the DM abundance slowly builds up from, say, the collision of the bath of (SM) particles. Although this mechanism is perfectly capable of explaining the present observed relic abundance of DM, but due to its super weak coupling with the visible sector, usually it is challenging to test such a framework in laboratory experiments or astrophysical observations. In a minimal $U(1)_X$ gauge extended SM that incorporates a vector boson DM which receives mass due to spontaneous symmetry breaking of $U(1)_X$ via radiative corrections known as the Coleman-Weinberg mechanism, we showed a mass scale is {\it transmuted} from the dark to the electroweak sector through the portal interactions. The underlying scale invariance allows only two independent parameters, namely, the DM mass $m_X\gtrsim 240$ GeV and the gauge coupling $g_X$. The new scalar is also found to be naturally light with mass $m_{h_2}\ll m_X$. We have shown that this set-up leads to DM production via freeze-in for suitable choice of interaction strength, that is within the reach of several low energy experiments. Considering $g_X\sim\mathcal{O}\left(10^{-5}\right)$ it is possible to produce the entire observed DM relic abundance for $m_X\sim\mathcal{O}\left(1~\rm  TeV\right)$ via 2-to-2 scattering of the SM particles in the thermal bath. Due to the presence of the light scalar mediator (with mass $\sim$ MeV), the parameter space for freeze-in DM can be constrained from spin-independent direct detection exclusion limits from experiments like PandaX-4T
%\ag{kindly name experiments and quote some numerical values}
, satisfying bounds from 
big bang nucleosynthesis (BBN), ensuring that the light scalar decays into SM degrees of freedom with a lifetime of less than a sec. For $m_{h_2}\sim\mathcal{O}\left(1~\rm GeV\right)$, one has to choose a very heavy DM that automatically satisfies direct detection bound.

%{\bb{add DD-intensity frontier complementarity, Fig.7}}

The main findings of this paper are summarized in Figs.~\ref{fig:exptLim1} \& \ref{fig:summaryPlt}. In the former, the requirement of obtaining the right abundance for the DM essentially fixes the scalar mixing to a fixed value of $\sin\theta\sim\mathcal{O}\left(10^{-5}\right)$, which is again attributed to the the scale invariance of the theory. This gives rise to a simple straight line contour in the $m_{h_2}-\sin \theta$ plane satisfying relic density and direct detection bounds. But interestingly this particular mixing is within the reach of several (proposed) lifetime and intensity frontier search facilities for $m_{h_2}$ up to about 3.5 GeV. In Fig. \ref{fig:summaryPlt}, we show the complementarity probes of such DM candidates, which tells us for massive DM $(m_X\gtrsim 1.8~\text{TeV})$ the intensity frontier experiments can provide more sensitivity than direct detection, even in probing parameter space that goes beyond the $\nu$-floor. As a consequence, even though the present model may produce null results in the scattering experiments but still leaves a window to be probed in experimental frontiers.
The scale invariant $\text{SM}\times U(1)_X$ symmetry thus leads to a testable freeze-in scenario satisfying all relevant constraints on the DM parameter space, where the coupling responsible for freeze-in is being directly probed at laboratories around the world.

%\ag{Comment on why mediator mass smaller than DM mass and leading to straight line.}{\bb{please mention this}}

We would finally like to conclude by pointing out that given such relevance of scale invariance and the freeze-in mechanism of DM formation in early universe along with the availability of planned and upcoming several light dark sector (intensity frontier and lifetime frontier) experiments, we showed that due to constrained relations between the model parameters ($ \lambda_{HS}\propto g_X^4 $) it uniquely predicts the FIMP portal coupling at which the relic density can be satisfied as well as be searched for in the laboratory including DM direct detection experiments which otherwise is not motivated for FIMP hunting. This ushers in a new era where UV-completion in BSM particle physics model-buildings may lead to predictive FIMP dark matter candidates to be tested in very near future.

\section{Acknowledgement}
%%%%%%%%%%%%%%%%%%%%%%%%%%%
The authors would like to thank Alexander Pukhov and Ahmad Mohamadnejad for helping in implementing the model in LanHEP and in micrOmegas~\cite{Belanger:2010pz}, also Nicolás Bernal, Sudip Jana and Nobuchika Okada for useful discussions. The authors would like to thank Dmitry Gorbunov for pointing out a mistake in PS191 limit. BB received funding from the Patrimonio Autónomo - Fondo Nacional de Financiamiento para la Ciencia, la Tecnología y la Innovación Francisco José de Caldas (MinCiencias - Colombia) grant 80740-465-2020. 
This project has received funding /support from the European Union's Horizon 2020 research and innovation programme under the Marie Skłodowska-Curie grant agreement No 860881-HIDDeN.
%%%%%%%%%%%%
\appendix
%%%%%%%%%%%%
\section{Annihilation cross-sections for freeze-in production}\label{sec:app-ann}
%%%%%%%%%%%%%%%%%%%%%%%%%%%%

Here we note down the analytical expressions for all the relevant 2-to-2 annihilation cross-section of the SM particles to DM final states. We denote all SM leptons by $\ell$, quarks by $q$ and gauge boson masses by $m_V\in m_W,m_Z$. The total decay width of the SM-like Higgs is indicated by $\Gamma_{h_1}\simeq 4$ MeV~\cite{CMS:2016dhk}.

%\left(s-m_{h_1}^2\right)

%\left(s-m_{h_1}^2\right)^2+\Gamma_{h_1}^2\,m_{h_1}^2

\begin{equation}\begin{aligned}
&\sigma\left(s\right)_{\ell\ell\to XX}\simeq \frac{g_X^4\,m_\ell^2}{64\pi\,s}\,\frac{\sqrt{\left(s-4m_X^2\right)\,\left(s-4m_\ell^2\right)}}{\left(s-m_{h_1}^2\right)^2+\Gamma_{h_1}^2\,m_{h_1}^2}\Biggl(\frac{m_{h_1}^2-m_{h_2}^2}{s-m_{h_2}^2}\Biggr)^2\Biggl(\frac{s^2-4m_X^2s+12m_X^4}{\left(m_X^2+g_X^2v_h^2\right)^2}\Biggr)
\\&
\sigma\left(s\right)_{qq\to XX}\simeq\frac{g_X^4\,m_q^2}{192\pi\,s}\frac{\sqrt{\left(s-4m_X^2\right)\,\left(s-4m_q^2\right)}}{\left(s-m_{h_1}^2\right)^2+\Gamma_{h_1}^2\,m_{h_1}^2}\Biggl(\frac{m_{h_1}^2-m_{h_2}^2}{s-m_{h_2}^2}\Biggr)^2\Biggl(\frac{s^2-4m_X^2s+12m_X^4}{\left(m_X^2+g_X^2v_h^2\right)^2}\Biggr)
\\&
\sigma\left(s\right)_{VV\to XX}\simeq\frac{g_X^4}{288\pi\,s}\,\sqrt{\frac{s-4m_X^2}{s-4m_V^2}}\,\Biggl(\frac{m_{h_1}^2-m_{h_2}^2}{s-m_{h_2}^2}\Biggr)^2\,\Biggl(\frac{1}{\left(s-m_{h_1}^2\right)^2+\Gamma_{h_1}^2\,m_{h_1}^2}\Biggr)\\&\Biggl[\frac{\left(s^2-4m_X^2s+12m_X^4\right)\left(s^2-4m_V^2s+12m_V^4\right)}{\left(m_X^2+g_X^2v_h^2\right)^2}\Biggr]
\\&
\sigma\left(s\right)_{h_1 h_1\to XX}\simeq \frac{g_X^4}{32\pi s}\Biggl(\frac{m_{h_1}}{m_X}\Biggr)^4 \frac{\left(s+2m_{h_1}^2\right)^2}{\left(s-m_{h_1}^2\right)^2+\Gamma_{h_1}^2\,m_{h_1}^2}\,\sqrt{\frac{s-4m_X^2}{s-4m_{h_1}^2}}\,\Biggl(1-\frac{4m_X^2}{s}+\frac{12m_X^4}{s^2}\Biggr).
\end{aligned}    
\end{equation}

\noindent The last expression is derived by keeping only the leading order in the double expansion of $g_X$ and $m_{h_2}^2/m_X^2$.

%%%%%%%%%%%%%%%%%%%%%%%%%%%%%%%%%
\bibliography{Bibliography}
%%%%%%%%%%%%%%%%%%%%%%%%%%%%%%%%%

\end{document}